\documentclass[12pt,preprint]{aastex}

\newcommand {\hii}{H {\small II}~}
\newcommand {\ha}{H$\alpha$~}
\newcommand {\flux}{ergs^{-1}cm^{-2}~}
\newcommand {\lum}{ergs^{-1}}
\newcommand{\kms}{\ifmmode {\rm \ km \ s^{-1}} \else $\rm \,km~ s^{-1}~$\fi}

\newcommand{\etal}{{\it et al.}~}

\begin{document}


\title{\ha Imaging of Early-type (Sa-Sab) Spiral Galaxies II: Global Properties \footnote{Based on observations obtained with the 
3.5-meter telescope at Apache Point Observatory (APO) and the 0.9-meter telescope at Kitt Peak National  
Observatory (KPNO). The APO 3.5m telescope is owned and operated by the 
Astrophysical Research Consortium.}}


\author{Salman Hameed\altaffilmark{2}}
\affil{Five College Astronomy Department, Smith College, Northampton, MA 01063}
\altaffiltext{2}{Visiting Astronomer, Kitt Peak National Observatory (KPNO).
KPNO is operated by the Association of Universities for Research in Astronomy, Inc. (AURA)
under cooperative agreement with the National Science Foundation.}
\email{shameed@ast.smith.edu}
\author{Nick Devereux}
\affil{Department of Physics, Embry-Riddle Aeronautical University, Prescott, AZ 86301}
\email{devereux@erau.edu}



\begin{abstract}
New results, based on one of the most comprehensive \ha imaging surveys of nearby Sa-Sab spirals completed to 
date, reveals early-type spirals to be a diverse group of galaxies that span 
a wide range in massive star formation rates. While the majority of Sa-Sab galaxies in our
sample are forming stars at a 
modest rate, a significant fraction ($\sim29\%$) exhibit star formation 
rates greater than 1 M$_{\odot}yr^{-1}$, rivaling the most prolifically star forming late-type
spirals. A similar diversity is apparent in the star formation history of Sa-Sab spirals as measured by their 
\ha equivalent 
widths. Consistent with our preliminary results presented in the first paper in this series, we find  giant \hii regions 
(L(\ha) $\ge$ 10$^{39}~\lum$) in the disks of $\sim~37\%$ of early-type spirals. We suspect that 
recent minor mergers or past interactions are responsible for the elevated levels of \ha emission and perhaps, 
for the presence of giant \hii regions in these galaxies. Our results, however, are not in total agreement 
with the \ha study of Kennicutt \& Kent who did not find any early-type spirals with  \ha equivalent 
widths $>$14\AA.  A close examination of the morphological classification of galaxies, however,
suggests that systematic differences between 
the Revised Shapley Ames catalog and the Second Reference Catalog may be responsible for the contrasting
results. 

\end{abstract}




\section{Introduction}

The last few decades have seen enormous progress in the understanding of
star formation in spiral galaxies. While there is good agreement concerning the wide range
of star formation rates seen 
among late-type spirals, different studies have yielded conflicting results for early-type (Sa-Sab)
spiral galaxies.  
The \ha equivalent width measurements of \citet{KK1983} have been the most influential in 
establishing a dependence of massive star formation rates on spiral Hubble type, with 
early-type spirals emerging as galaxies with preponderantly low star formation rates,
especially when compared to their later-type counterparts.

Other \ha studies, however, find evidence for copious star formation in a significant fraction of the early-type spiral 
population. \citet{Young1996}, \citet{Usui1998} and  \citet{HD1999} have identified numerous 
field early-type spirals with massive star formation rates comparable to Sc galaxies. The presence 
of such prolifically star forming early-type spirals weakens or completely removes the dependence of star formation rates
along the spiral Hubble sequence. 

The disparity of results regarding star  formation rates in early-types spirals extends to 
other star formation indicators. For example, analyses of the far-infrared measurements \citep{DY1990, Tomita1996,DH1997} 
were unable to reveal any dependence of massive star formation rates 
on spiral Hubble type. On the contrary, they were able to identify specific Sa-Sab galaxies with prodigious 
massive star formation as measured by their far-infrared emission. In contrast, \citet{RL1986} 
and \citet{PR1989} also analyzed the far-infrared data and 
found star formation rates to be inhibited in early-type spirals.

We have launched a systematic \ha survey of early-type spirals in the local universe to 
quantify and analyze the properties of ionized gas in these galaxies in an effort
to reconcile the conflicting results from 
earlier studies. Preliminary results, based on 27 galaxies, were presented in 
Hameed \& Devereux (1999; hereafter HD99). Here, we present the results of the entire sample.
The sample is defined in $\S$2 and the observations are described in $\S$3. Our results are presented
in $\S$4, followed by a discussion in $\S$5. Our conclusions are stated in $\S$6.

\section{The Sample}
The galaxies have been selected from the Nearby Galaxy 
Catalog (NBG) \citep{Tully1988}, which includes, among other galaxies, a list of all (74) {\it known} 
early-type spirals that are brighter than 12.0 magnitude and located within 40 Mpc. 
Our survey is motivated by the Infrared Astronomical Satellite (IRAS) survey 
that covered roughly 96$\%$ of the sky. The original goal of our survey was to image the 57, noninteracting, 
early-type (Sa-Sab) spirals that were also scanned by IRAS, that are brighter than m(B)=12.1, and have 
velocities less than $3000 \kms$, placing them nearer than 40 Mpc (H$_{0}$
= 75 km s$^{-1}$ Mpc$^{-1}$). All of these early-type spirals, with the exception of NGC 7727, were detected by the 
IRAS satellite. 

\ha images for 27 galaxies were presented in HD99, six of which have m(B) $>$ 12.1 
but were included because of their unusually high 
far-infrared luminosities (see HD99 for details). We have observed 18 additional early-type spirals, 
and have obtained \ha images for six more galaxies from other sources. Thus, {\it the results presented 
in this paper are based on a sample of 51 early-type spirals},  45 out of 57 ($\sim79\%$) of which have m(B)$\le$12.1. Table 1 lists some useful observables for the 
target galaxies. Because of the limited resolution of IRAS satellite, our study excluded other 
spirals with cataloged 
companions within  $3\arcmin$ of each other, thus eliminating almost all early-type spirals that
are strongly interacting. 
While we would have liked to include all early-type spirals, ours constitutes essentially a complete 
sample and the \ha images of 51 spirals presented herein
should provide a representative picture of Sa-Sab galaxies in the local universe.

\section{Observations}
\subsection{New \ha Observations}
Fourteen early-type spirals were observed using the direct CCD Imager on the 
0.9 m telescope located at Kitt Peak National Observatory (KPNO) in Arizona. The imager 
used a 2048 $\times$ 2048 Tektronics chip and has a pixel scale of $0\arcsec.68~$pixel$^{-1}$
at f/7.5, yielding a field of view of $23\arcmin.2 \times 23\arcmin.2$. All fourteen galaxies 
were imaged using the 72\AA, narrow-band, \ha + [NII] (hereafter \ha, unless otherwise noted)
filter centered at 6586\AA. 
The off-band images were obtained using the narrow-band line-free filter 
centered at 6487\AA ($\Delta \lambda~=~67$\AA). Three exposures of 1200 s were 
obtained through each of the line and continuum filters. Details of the observations 
are summarized in Table 2.

Three galaxies were imaged using the Seaver Prototype Imaging camera (SPIcam) on the 
Astrophysical Research Consortium (ARC) 3.5 m telescope at Apache Point 
Observatory (APO) in New Mexico. SPIcam uses a 2048 $\times$ 2048 CCD, 
has a pixel scale of $0\arcsec.14$ pixel$^{-1}$, and a $4\arcmin.8$ field 
of view. The pixels were binned 2 $\times$ 2 in the readout, resulting in a pixel scale of 
$0\arcsec.28$ pixel$^{-1}$. A 70\AA~\ha filter, centered at 6610\AA, and 
a 120\AA~line-free continuum filter, centered at 6450\AA, were used to obtain 
the line and continuum images, respectively. Three exposures of 300 s were obtained through each of
the line and continuum filters (Table 2).  

NGC 2985 was observed using the Double Imaging Spectrograph (DIS) with the 
3.5 m telescope at APO. DIS has a pixel scale of $0\arcsec.61$ pixel$^{-1}$ and 
a $4\arcmin.2$ field of view. Three 390 s exposures were obtained through 
each of the line (6610\AA, $\Delta \lambda~=~70$\AA) and the continuum 
(6450\AA, $\Delta \lambda~=~120$\AA) filters (Table 2). 

Observations for all galaxies were obtained under photometric conditions. The 
\ha images were flux-calibrated using the standard 
stars BD $+28^{\circ}4211$, PG 0934+554, Feige 34, and HZ 44 \citep{Massey1988}. The data was reduced
in the same manner as explained in HD99.

\subsection{\ha Data from Other Sources}
Twenty-one  galaxies from HD99 were observed with the 1.5 m 
telescope located at Cerro Tololo Inter-American Observatory (CTIO) in Chile. \ha 
images for six galaxies were obtained with the ARC 3.5 m telescope 
located at Apache Point Observatory. Details of these observations
are given in HD99.

NGC 3031 (M 81) was imaged by \citet{Devereux1995} with 
the Case Western Burrell Schmidt telescope at Kitt Peak National Observatory. 
The \ha + [NII] image was obtained with a narrow-band, 74\AA, filter centered at 
6568\AA. A narrow-band, 72\AA, line-free  filter centered at 6481\AA, 
was used to obtain the continuum image. Results and analysis of the M 81 
data are presented in \citet{Devereux1995}.

Images for five galaxies (NGC 3705, NGC 4192, NGC 4419, NGC 4450, and NGC 4984)
were kindly provided by Rebecca Koopman. All of these galaxies, except NGC 4984, were imaged
with the 0.9 m telescope at KPNO. The \ha image of NGC 4984 was obtained with the 
0.9 m telescope located at CTIO. All of Koopman's images were obtained with a narrow-band 
\ha filter and a broadband R filter was used for the continuum images. 
Details of these observations are presented in \citet{Koopman1997}.   

\subsection{Comparison with Previous Studies}
There are several galaxies in our survey that have published \ha fluxes. Figure 1 
compares \ha fluxes for 15 galaxies measured by us in the same aperture as 
those in the literature. Figure 1 includes fluxes for seven galaxies published previously 
in HD99. Overall, there is good agreement between the measurements.
There are, however, two exceptions. Our \ha flux for NGC 4736 is 196$\%$ below the 
value obtained by \citet{Young1996} and our \ha measurement for NGC 3718 is 
180$\%$ above the value quoted by Young \etal (1996). \citet{KK1983} also measured the \ha flux of NGC 4736 and their value is within
29$\%$ of our measured flux. We do not know the reason for the difference 
of our flux value from that of Young \etal. However, we should note that there are 4 other  galaxies 
that are in common with  Young \etal's sample, and their fluxes are within 2$\%$(NGC 3504),
5$\%$(NGC 660), 23$\%$(NGC 2146), and 67$\%$(NGC 3623) of our values. 
Overall, the mean ratio of our flux measurements to the other studies is 1.19 $\pm$ 0.6. However, 
the ratio drops down to 1.10 $\pm$ 0.4 if we exclude NGC 3718 and NGC 4736.

\subsection{Uncertainties in \ha Flux and Equivalent Width Measurements}

\subsubsection{[NII] Contamination}
The \ha fluxes and \ha equivalent widths presented in this paper include
contributions from the two satellite [NII] lines at 6548\AA~and 6584\AA.
Complete inclusion of these lines in our line filters  allows the possibility to later correct 
for [NII] contamination, when more information is available.
Previous work has indicated that the [NII]/\ha 
ratio varies from one galaxy to the next and within individual galaxies
\citep{KK1983, Kennicutt1992}. 
A constant factor can, in principle, be used to correct for [NII] contamination. However, 
galaxy to galaxy variations are large enough to compromise  this 
procedure. \citet{Kennicutt1992} examined [NII]/\ha ratios for 90 nearby galaxies, and found a median value
close to 0.53 (excluding Seyfert galaxies). However, the mean ratio for 6 non-interacting 
Sa-Sab galaxies in his sample was 1.24 with ratios ranging from 0.48 up-to 
as high as 2.4. Furthermore, within a particular 
galaxy, the diffuse ionized gas has a higher value of [NII]/\ha than \hii regions
\citep{Greenawalt1998}. The [NII]/\ha ratio is especially high in the central regions of galaxies, where 
\ha absorption is strongest and [NII] is in emission \citep{Young1996}.
Thus we would need precise information about [NII]/\ha ratio both within and among galaxies to 
properly correct for [NII] contamination. 
  
\subsubsection{Extinction} 
The \ha fluxes presented in this paper have not been corrected for Galactic or internal 
extinction. Thus, the \ha fluxes and luminosities we measure provide lower limits to the 
intrinsic \ha fluxes and luminosities, and consequently to massive star formation 
rates. \citet{KK1983} estimated, on average, 1 magnitude of extinction
in their \ha fluxes. However, extinction is expected to be higher for galaxies 
with high inclinations and for galaxies with dusty starbursts. Detailed studies of the 
highly disturbed early-type spirals, NGC 2146 and NGC 660, estimate 9 and 13 magnitudes of 
extinction in the visible, respectively \citep{Young1988}. In such extreme  cases, Balmer 
recombination lines in the infrared, like Paschen $\alpha$ and Brackett $\gamma$,
will be  more suitable for determining star formation rates.

\subsubsection{Continuum Subtraction}
The continuum image for each galaxy was scaled to the line plus continuum 
image by measuring the integrated fluxes of 10-15 foreground stars common to both images. 
This scale factor, however, often needs adjustments, since the foreground stars 
and the galaxy are sometimes different in color. Adjustments were made iteratively 
until a satisfactory subtraction was obtained for the majority of the galaxy. The application
of a constant scale factor across the entire galaxy introduces significant uncertainty, especially
if there are large variations in color caused by changes in stellar populations. Often, the central 
regions are over-subtracted when the disk is well fit. 

The uncertainty in the \ha flux depends sensitively and non-linearly on the continuum level 
subtracted. This uncertainty is further aggravated by the contribution of the large 
stellar bulge in early-type spirals. For most galaxies in our sample, a 2-5$\%$ variation 
in the continuum level results in 10-50$\%$ errors in the \ha flux, depending
on the relative contribution of the continuum light. Flux measurements for NGC 4594 
(The Sombrero galaxy) are particularly uncertain. 
This is an edge-on galaxy where the continuum light is completely dominated by the
bulge. In the absence of a good view of the disk, it is very difficult to ascertain 
the true continuum level. A $3\%$ variation in the continuum level of NGC 4594 results in a $\sim$130$\%$ 
change in the measured \ha flux.

\subsubsection{Measurement of \ha equivalent widths}
\ha equivalent widths are usually measured spectroscopically, however, images
may also be used. The \ha equivalent widths presented 
in this paper were calculated by dividing the continuum-subtracted \ha emission line 
fluxes by the associated continuum flux and expressing the result in 
Angstroms (\AA) after the appropriate unit conversions. Unlike integrated spectroscopy, images provide 
additional information about the distribution of \ha emission within a galaxy. 
For example, almost the entire \ha emission of NGC 4369 is concentrated 
in the nuclear region. When measuring the continuum flux, we have included 
the full extent of the galaxy, which results in an \ha equivalent width of 16.5\AA.
However, if we use the continuum flux covering the same region as the \ha emission,
the value of \ha equivalent width rises to 20.5\AA. In order to be 
consistent with existing spectroscopic measurements, we have used global \ha and continuum 
fluxes for the determination of \ha equivalent widths in this paper. However, we note 
in passing that this problem does not exist for 
late-type spirals where the massive star formation is usually spread throughout the entire
disk and, as a result, is coextensive with the past star formation.

\section{Results} 
\subsection{Category 1 and Category 2 Early-type Spirals}
The \ha fluxes for 51 early-type spirals are presented in Table 3, along with
the calculated \ha luminosities and \ha equivalent widths. In order to understand
the diverse nature of early-type spirals, we have followed the division of 
Sa-Sab galaxies by HD99, based on the luminosity of the largest \hii region 
in the disk of the galaxy. The motivation for this division comes from the studies of 
\citet{KEH1989} and \citet{Caldwell1991}, which did not find any \hii regions with 
 \ha luminosity $\ge 10^{39}~\lum$ in their samples of Sab and Sa galaxies, respectively.
Thus, in our study, all \hii regions in the disk of Category 1 early-type spirals have 
L(\ha)$< 10^{39}~\lum$ whereas, Category 2 galaxies host  at least one 
\hii region {\it in the disk} with \ha luminosity $\ge 10^{39}~\lum$. 
We have also identified galaxies with intense nuclear starbursts (NGC 1022, NGC 1482, 
NGC 3885, NGC 3471) and have included them as Category 1 galaxies because
they have virtually no disk \ha emission. \ha images for 
Category 1 and Category 2 galaxies are presented in Figures 2 and 3, respectively.

Figures 4 and 5  show the range of global \ha luminosities and \ha equivalent widths for the two categories of early-type spirals. 
As expected, Category 1 galaxies have preferentially lower global \ha luminosities and smaller \ha equivalent widths compared 
to Category 2 galaxies. The luminosity of the prototypical early-type spiral, M 81, 
is also marked on these figures. Despite some overlap, a two-tailed 
Kolmogorov-Smirnov (K-S) test indicates that the two categories are
not derived from the same population at a confidence level greater than 99$\%$. 

Within the current sample of 51 nearby galaxies, we find that 59$\%$ belong to Category 1. However, 
a significant fraction (37$\%$) of early-type spirals, host giant \hii regions in their disks. Two galaxies,
NGC 660 and NGC 2146, have highly disturbed morphologies and have not been 
classified into either of the categories (HD99).

Despite having similar optical morphologies, early-type spirals show a wide diversity
in H$\alpha$ morphology. Most Category 1 galaxies appear undisturbed in the continuum image, but exhibit
diversity when it comes to the nuclear H$\alpha$ emission. Almost half (14/30) of all Category 1 galaxies host 
Extended Nuclear Emission-line Regions (ENER); a region of diffuse ionized gas
in the nuclear region. As has been noted by \citet{Keel1983}, ENERs are only visible when there is very little or no 
star formation near the nucleus. In seven other Category 1 galaxies, its hard to detect ENERs due to 
their high inclination. Two Category 2 galaxies 
(NGC 3169 and NGC 7213) also exhibit this diffuse gas. Seyfert nuclei have been 
identified in four Category 1 early-type spirals.

In HD99 we had speculated on a possible direct correspondence between the spectroscopic
classification of 
LINERS and the morphologically identified ENERs. However, due to the small number of galaxies
with spectroscopic classifications, we could not address that assertion
statistically in that paper. Now, with a sample of 51 galaxies, we find that 12 Category 1 galaxies 
have been classified as LINERs, and 10 of those galaxies also show ENER emission, {\it suggesting an 
almost one-to-one correspondence between the spectroscopic and morphological classifications.} 

The nuclei of Category 2 galaxies have been mostly classified as 'H' spectroscopically, indicating the 
presence of \hii regions (Table 4). There are, however, four Category 2 galaxies with Seyfert nuclei. {\it The presence of 
equal numbers of Seyfert nuclei in each of the categories suggests that Seyfert activity is unrelated 
to the global star forming properties of early-type spiral galaxies.}

While the division of early-type spirals into two categories has been useful, we have 
to be cautious. First, it is difficult to measure the \ha 
flux of an individual \hii region. In HD99, we measured \ha fluxes for the largest \hii regions
manually, using circular apertures. However, since then we have been using an 
automated program, ``HIIphot'' (See \citet{Thilker2000} for details of the program), to determine 
the luminosity functions for star forming regions in spiral galaxies (Hameed, Thilker, \& Devereux 
{\it in preparation}). All of our manual classifications are consistent with HIIphot, with the 
exception of NGC 3169 and NGC 7213. Both of these galaxies host \hii regions with 
L(\ha)$> 10^{39}~\lum$, and thus have been classified as Category 2 galaxies in the 
current paper. 

Second, we do not yet know of any physical significance of our classification scheme. A few studies, however,
 have suggested that the physical properties of \hii regions change between $10^{38}-10^{39}~\lum$, such as
the transition from normal ``giant'' \hii regions to ``supergiant'' \hii regions \citep{KEH1989}, the
transition from sparse embedded star clusters to dense embedded clusters in individual \hii 
regions \citep{OC1998}, and the claim of the change from radiation bounded to density 
bounded \hii regions \citep{Beckman2000}.   
Whatever the underlying significance of our scheme, it is of sufficient importance to report the detection of a significant 
fraction of early type spirals with giant \hii regions, which is in contrast to the earlier 
findings of \citet{KEH1989} and \citet{Caldwell1991}.

\subsection{\ha vs. Far-infrared}
The presence of massive stars can be traced via both \ha and far-infrared emission.
The relationship between the global \ha and far-infrared (40-120 $\mu$m) luminosities 
is presented in Figure 6 for 51 Sa-Sab galaxies. Open dots represent Category 1 
early-type spirals, Category 2 galaxies are represented by solid dots, and the two unclassified galaxies, 
NGC 660 and NGC 2146, have been plotted as stars. The far-infrared
luminosity, L(FIR), has been calculated using,
$$L(40-120 \mu m) = 3.65 \times 10^{5}~ S(FIR) ~D^{2} ~L_{\odot}$$
where S(FIR) = (2.58 S$_{60\mu m}$ + S$_{100\mu m}$) and D is the distance expressed in Mpc. 
 The values S$_{60\mu m}$ and S$_{100\mu m}$
are  fluxes measured by the Infrared Astronomical Satellite (IRAS) in units of Jy. The solid lines in 
Figure 6 identify ratios of L(FIR)/L(\ha) expected for \hii regions, ionized by massive stars
ranging in spectral type from O5 to B5. The ratios are based on a simple model of 
an idealized \hii region, in which all photons shortward of the Lyman limit  are 
absorbed by the hydrogen gas while the bolometric luminosity of the star is absorbed by surrounding dust and 
re-radiated in the far-infrared \citep{Panagia1973, DY1990}. 

Figure 6 shows that most early-type spirals in our sample, including M 81 (NGC 3031),
have L(FIR)/L(\ha) ratios consistent with those expected for \hii regions. However, 
there are some galaxies that have unusual ratios.
Most of the errant galaxies either host a Seyfert nucleus (NGC 4151, NGC 2273, 
NGC 7172) or are nuclear starbursts (NGC 1022, NGC 1482).  
Non-stellar emission from Seyfert galaxies is not expected to follow the 
L(FIR)/L(\ha) ratios for massive stars, whereas, the unusual location of nuclear starburst
galaxies in Figure 6 may be due to high dust extinction. Similarly, the locations of
NGC 660 and NGC 2146 may be explained by the 13 and 9 magnitudes of extinction, respectively
\citep{Young1988}. The nucleus of NGC 5188 is spectroscopically classified as ``H'' by \citet{VV1986}. However, we
detect an unresolved \ha nuclear point source (HD1999) and suspect that its has a Seyfert nucleus that may have been missed
in the earlier spectroscopic study.  The remaining 4 outlying galaxies (NGC 1371, NGC 3718,
NGC 4419, NGC 4845) warrant additional scrutiny to fully understand 
the reason for their unusual ratios.

Another word of caution here. Despite the fact that massive stars appear to be capable of 
sustaining the \ha and far-infrared luminosities in most galaxies, detailed studies of the nearest 
spiral galaxies have indicated that a sizable fraction of the bulge far-infrared and 
\ha emission is not powered by massive stars at all \citep{Devereux1994, Devereux1995}. Approximately 
30$\%$ of the total far-infrared and \ha emission from M 31 and M 81 originates 
from the bulge. This bulge \ha emission is diffuse in nature, much like the ENERs, and observations 
with the Hubble Space Telescope have convincingly shown the absence of massive stars in the central regions of 
these two galaxies \citep{Devereux1997}. As mentioned in $\S$ 4.1, almost half of all 
Category 1 early-type spirals host this diffuse \ha emission in the nuclear region.
The source of the \ha emission is unknown, however, it is unlikely to be ionized by a central AGN in all cases, as
the luminosity of the extended region far exceeds that of the nucleus itself in several galaxies. On the other
hand an extended region of ionization could be sustained by 
bulge post-asymptotic giant branch (PAGB) stars or by collision induced shocks occuring within the bulge
gas itself \citep{Heckman1996}. The same bulge PAGB stars may also be responsible for heating the dust that gives rise
to the diffuse bulge far-infrared emission.

Figure 7 investigates the correlation between L(FIR)/L(B) and \ha equivalent widths
for all 51 galaxies in our sample. Both quantities measure the ratio of present to 
past star formation, where the present star formation rate is determined by far-infrared and
\ha emissions and are normalized by past star formation as traced by blue and red continuum luminosities, 
respectively. Figure 7 shows that, overall, there is a trend between 
\ha equivalent width and L(FIR)/L(B), but its not a one-to-one correlation. 
In general, galaxies with low L(FIR)/L(B) have low \ha equivalent widths, and 
galaxies with high values of L(FIR)/L(B) have high \ha equivalent widths.
A Spearman's Rank Correlation test gives a value of 0.68, suggesting 
a weak correlation. The least-square fitting to the data gives a slope 
of 0.47. Note that some  of the galaxies with excess L(FIR)/L(B)
are also the outliers in Figure 6 and vice versa. Heavy extinction may be responsible for the 
shallow slope of the relation. \citet{Usui1998}, based on a study of 15 galaxies,  find a slope of 0.82 for 
L(FIR)/L(B) versus \ha equivalent widths for early-type spirals with high EW(\ha) $>$ 3\AA, and a correlation 
coefficient 
of 0.79. Our results do not change significantly even after restricting our sample  to 
galaxies with  EW(\ha) $>$ 3\AA.

\subsection{\ha Equivalent width vs. (B-V)}
The relationship between the \ha equivalent widths and (B-V)$_{e}$ colors 
for 49 early-type spirals is illustrated in Figure 8. (B-V)$_{e}$ colors are
adopted from the Third Revised Catalog of Bright Galaxies \citep{deVauc1991}.
NGC 4750 and UGC 3580 do not have (B-V)$_{e}$ measurements, and hence are 
not included here. Figure 8 shows that early-type spirals exhibit a wide range of 
\ha equivalent widths and  (B-V)$_{e}$ colors, much wider than previously reported for early-type 
spirals by \citet{Kennicutt1983}. Furthermore, the \ha equivalent widths and (B-V)$_{e}$
colors are correlated. A non-parametric Spearman rank correlation test allows 
the null hypothesis, that the two observables are not correlated, to be 
rejected at $>$ 99.9$\%$ significance level.

As noted by \citet{Kennicutt1983}, the relationship between the \ha equivalent widths
and (B-V) colors can, in principle, provide useful constraints on the past star formation
history. By way of illustration, the range of plausible star formation histories permitted by 
the new observations is also indicated in Figure 8. The solid lines in Figure 8 illustrate the 
loci of model \ha equivalent widths and (B-V)$_{e}$ colors expected for a 15 billion year old galaxy 
that has evolved at different rates. There are two curves plotted, each running from upper left to lower right. 
The left curve represents zero extinction, the right curve includes the extinction expected for 
a galaxy viewed essentially edge-on (i = 85$^\circ$). The flatter dashed lines joining points on 
the two curves show the direction of reddening vectors. The results are based on a 
spectrophotometric
evolution model for starbursts and galaxies called PEGASE \citep{FR1997, Moy2001}.

The details of the PEGASE evolutionary models are presented in \citet{FR1997, Moy2001}. 
Briefly, the model evolves stars with a range of 
stellar metallicities from the main sequence, to the He flash, the horizontal branch, the 
asymptotic giant branch, and finally to demise as supernova or white dwarfs. The star 
formation scenarios used here are similar to those adopted by \citet{Kennicutt1983}; an exponential 
star formation law with an ``extended'' Miller-Scalo initial mass function, and stellar masses
ranging from 0.1 - 100 M$_{\odot}$. PEGASE calculates broad band colors as well as 
nebular emission. Additionally, radiative transfer computations provide an internally
consistent estimate of dust extinction in the model galaxy.  

The blue color and high \ha equivalent width combination at the upper left of Figure 8
denotes a galaxy that has been converting gas into stars at essentially a constant rate 
since birth. The reddest colors and smallest \ha equivalent widths at the lower 
right of the diagram denotes a star formation history in which most of the 
primordial gas was rapidly converted into stars some 3 $\times~ 10^{10}$ years
after galaxy formation. Most early-type spirals lie between the two extremes described above. Thus, the 
PEGASE results indicate that the star formation history of early-type spirals
is diverse; a further testament to their heterogeneous nature already apparent 
from the \ha images. It is quite remarkable indeed that a history of continuous
star formation, or a continuously decreasing star formation rate, together with the 
effects of reddening, can explain the wide range of colors and \ha equivalent 
widths observed for the majority of  early-type spirals plotted in Figure 8.

Interestingly, the correlation between the \ha equivalent widths and (B-V)$_e$
colors breaks down for the objects with the highest \ha equivalent widths (i.e. $>$ 14\AA),
a result that is substantiated by a Spearman rank correlation test. Part of the reason
may be attributed to the presence of active galactic nuclei (AGNs). NGC 4151, for example,
is a well-known Seyfert 1 galaxy, and NGC 2273 and NGC 6810 are Seyfert 2's 
\citep{HFS1997, VV1986}. However, not all AGNs exhibit 
unusual colors in Figure 8. Conversely, not all of the errant points 
in Figure 8 can be attributed to AGNs. NGC 1482 and NGC 972, for example, have no 
documented evidence for an AGN and yet they are among the most deviant of all 
the outliers in Figure 8. The \ha images show evidence for a nuclear starburst in 
NGC 1482, and extended nuclear star formation activity in NGC 972. The 
unusually high \ha equivalent widths, given the red colors of NGC 1482 and 
NGC 972, are in fact, exactly what one would expect for a recent starburst 
superimposed on an older population. The high \ha equivalent 
widths indicate that the starburst increased the star formation activity by at least 
one order of magnitude over the pre-existing rate. Interestingly, recent HI observations of 
NGC 972 reveal a tidal tail in neutral hydrogen indicative of a past
interaction \citep{HY2003}. Note, however, that {\it not all} interacting early-type spirals 
have unusual colors. For example, M81 \citep{Yun1994}, NGC 3471 \citep{HY2003}, 
NGC 3885 \citep{HY2003}, NGC 4725 \citep{Wevers1984} show interaction signatures in HI 
and yet have colors and \ha equivalent widths that can be explained by a "normal" star formation 
history.

We caution that PEGASE interprets global measurements, which do not 
constrain or preclude different star formation histories {\it within}
individual galaxies. Nevertheless, the star formation histories 
of early-type spirals as revealed by their global (B-V)$_e$ colors 
and \ha equivalent widths are considerably more diverse than had
previously been appreciated (e.g. \citet{Kennicutt1983}). We have also included Kennicutt's Sa-Sab 
galaxies in Figure 8, most of which have preferentially smaller \ha equivalent widths and redder colors. 
We will explore some of the reasons for these differences in $\S$ 5.

\subsection{Massive Star Formation Rates}
The ionizing flux from galaxies can be converted to star formation rates using evolutionary synthesis models
and by making assumptions regarding abundances and the shape of the initial mass function (IMF) \citep{KTC1994, Kennicutt1998}. 
While we can, under some circumstances, measure low mass star formation in nearby molecular clouds in our own Galaxy, 
such measurements are impossible for external 
galaxies. Nevertheless, we have calculated total (low mass plus high mass) star formation rates using 
Kenniucutt's (1998) conversion: 
$$SFR(M_{\odot}yr^{-1})=7.9~\times~10^{-42}L(H\alpha)(ergs~s^{-1})$$
assuming solar abundances, a Salpeter IMF (0.1-100 M$_{\odot}$), and a Case B recombination at
T$_{e}$=10,000 K.  We find that the average star formation rate for our entire sample is 0.94 M$_{\odot} yr^{-1}$ 
with 29$\%$ of early-type spirals having star formation rates in excess of 1 M$_{\odot}yr^{-1}$, up to a 
maximum of 3.5 M$_{\odot}yr^{-1}$. 

While these calculations are interesting, they should be regarded with caution. Apart from the usual concerns 
regarding the shape and extent of the IMF, there are 
observational uncertainties that can affect the measured star formation rates. For example, we have
not corrected our \ha fluxes for extinction (see $\S$ 3.4.2) and yet numerous galaxies in our sample 
harbor starbursts in their nuclear regions or exhibit prominent dust lanes. The extinction in the optical can 
be as high as 13 magnitudes as seen in NGC 2146 \citep{Young1988}. Thus the calculated 
star formation rates certainly represent a lower limit, and in some cases they may be underestimated
by an order of magnitude. Second, the conversion factor used to the calculate star
formation rates assumes that almost all of the \ha emission is coming from \hii regions, but we have
discovered that ENER's are common in early-type spirals, consequently not all the \ha emission can be
attributed to massive star formation. While the above mentioned problems may be important 
in individual cases, these do not affect our overall conclusions regarding star formation rates in early-type spirals 
as none of these special circumstances ever dominate the global \ha emission of the galaxies, except in the case of NGC 4151, 
where the majority of the \ha emission is coming from the Seyfert nucleus.

\section{Discussion}

\subsection{The Ratio of Present to Past Star Formation Rates} 
Our comprehensive volume limited \ha survey has exposed 
the heterogeneous nature of nearby early-type spirals. Perhaps the most surprising 
result, first suggested in HD99, is the discovery that a significant fraction of Sa-Sab galaxies 
host giant \hii regions and levels of \ha emission comparable 
to those seen in nearby late-type spirals. 

The popular  perception that early-type spirals are currently associated with low levels of star formation is primarily due to 
the \ha equivalent width results of \citet{KK1983}. The star formation history of galaxies can be traced using \ha equivalent widths, where 
the current star formation, as measured by the \ha emission, is normalized by the red continuum light 
which represents the past star formation integrated over the lifetime of the galaxy. Figure 9 compares \ha equivalent 
widths of 51 Sa-Sab galaxies in our sample with those of Kennicutt \& Kent.  The top panel shows that 
early-type spirals exhibit a wide range of \ha equivalent 
widths, much wider than the Sa-Sab galaxies in the \citet{KK1983} sample
which have preferentially low \ha equivalent widths (second panel in Figure 9). Over half of Kennicut
\& Kent's  measurements are 
upper limits. Of the remaining detections, none exceed \ha equivalent widths of 
14\AA. In comparison, about 35$\%$ of early-type spirals in our sample have 
\ha equivalent widths $>$ 14\AA. 

The third panel of Figure 9 shows \ha equivalent widths for early-type spirals included in \citet{Usui1998}. However,
their study focused only on those Sa-Sab galaxies that have 
high L(FIR)/L(B) ratios. Consequently, one can expect the \ha 
equivalent widths in their sample to have preferentially higher values because of the
correlation in Figure 7 described previously in section 4.2.
Combining L(FIR)/L(B) luminosity ratios \citep{Tomita1996} and
\ha equivalent widths, Usui \etal estimated that 30$\%$ of early-type spirals 
must have \ha equivalent widths greater than 20\AA. We confirm their prediction, since we 
find 28$\%$ of Sa-Sab galaxies in our sample with \ha equivalent widths $>$ 20\AA. Figure 9 also includes \ha equivalent widths for 20 Sa-Sab galaxies taken from a recent survey of 334 galaxies across all Hubble types, conducted by \citet{James2004}.  While missing the very high star forming early-type spirals, the \citet{James2004} sample still display a range of \ha equivalent widths that is broader than \citet{KK1983}. {\it Overall, our survey of early-type spirals encompasses the range of \ha equivalent widths observed by \citet{KK1983}, \citet{Usui1998}, and \citet{James2004}}.

Seyfert galaxies were not included in the \citet{KK1983} and 
\citet{Usui1998} samples. The non-stellar emission from Seyfert nuclei can potentially
increase the \ha equivalent widths of galaxies, although it is really obvious only for one galaxy in
our sample; NGC 4151. 
Nevertheless, in order to remove any doubt over the influence of Seyferts, we have plotted our
\ha equivalent widths  again in the bottom panel of figure 9 after excluding the nine 
Seyferts in our sample. The overall trend of \ha equivalent 
widths is unaffected, suggesting that the presence of Seyfert nuclei does not dominate the global 
\ha emission of these early-type spiral galaxies. 

\subsection{Morphological Classifications: RC2 vs RSA}
There are at least two reasons for the differences between our results and those of 
\citet{KK1983}. First, a significant fraction  of the early-type spirals 
with high \ha equivalent widths reside in the southern hemisphere, and most galaxies
in Kennicutt \& Kent's sample are located in the northern hemisphere. The biggest factor, however,
may be the systematic difference in the morphological classification of galaxies
between the Revised Shapley Ames Catalog (RSA; \citet{ST1981}) used by \citet{KK1983}
and the Nearby Galaxy Catalog (NBG; \citet{Tully1988}) used by us. The NBG adopted most of
the morphological classifications from the Second Reference Catalog (RC2; \citet{devauc1976}). We have
listed the classifications from all three catalogs in Table 5 for the early-type spirals in our study.
There is, in general, good agreement among the different catalogs for Category 1 early-type spirals. 
For Category 2 galaxies, however, the RSA has systematically classified them 
as later-type galaxies. Figure 10 illustrates systematic
differences in classification between the RSA and RC2. Whereas, the misclassifications in Category 1
galaxies tend to be random, the misclassifications of Category 2 galaxies
are biased toward the later types. 

HD99 related the reason for these systematic differences to the 
subtle differences in the criteria used to classify spiral galaxies. 
The RC2 catalog uses all three Hubble criteria for classifying spiral galaxies:
the size of the bulge, the pitch angle, and the resolution of spiral arms \citep{devauc1959}. On the other hand, 
RSA classifications of spiral galaxies are based only on the characteristics 
of spiral arms \citep{Sandage1961}, which, in turn, are directly correlated with star formation. The
Carnegie Atlas of Galaxies \citep{SB1994} suffers from the same morphological bias: whereas the 
resolution of spiral arms in the Carnegie Atlas is replaced by a more useful parameter, the star 
formation rate, as determined by the \ha emission, the 
classification of Sa-Sab galaxies still follows those of the RSA catalog. This is surprising since the 
Carnegie Atlas, unlike the 
RSA catalog, proclaims to include bulge size as one of the classification criterion. However, a closer 
inspection of the two catalogs 
shows that the classifications of Sa and Sab galaxies are almost identical with only one exception: 
NGC 5121 changed from being an S0(4)/Sa in the RSA to an Sa in the Carnegie Atlas.    

Thus, we maintain that the wide variety of results in the published literature
concerning the systematics of massive star formation along the Hubble sequence can be traced directly to these 
subtle differences
in classification schemes between the RSA and the RC2. \citet{KK1983}, quite likely overlooked
several of the most prolifically star forming
early-type spirals because they were classified as later type spirals in the RSA catalog. Our view today of
massive star formation along the spiral Hubble sequence may have been quite different 
if previous investigators had exclusively adopted the RC2 classification scheme. 

\subsection{Interaction-induced Massive Star Formation?}
One of the most important results from our survey has been the discovery of a significant fraction of early-type spirals 
with giant \hii regions and star formation rates comparable to those found in late-type spirals.
Since the majority of Sa-Sab galaxies have more modest star formation rates, it is pertinent to ask if 
those with high star formation rates are going through a temporary phase of enhanced star formation? 

Studies over the past decade have exposed dynamical or morphological anomalies such as counter-rotating 
disks, shells, and disturbed bulges, in a number of early-type spirals. 
Such features represent the fossilized signatures of an interaction or a minor merger 
that occured in the distant past \citep{Schweizer1990, HS1992}. 
While most of these galaxies are quiescent today, they may well have experienced a 
temporary phase of enhanced star formation during a gravitational encounter with another galaxy. 

It is well established that  interacting galaxies exhibit higher levels of \ha \citep{Kennicutt1987} and FIR emission \citep{Bushouse1987} 
compared to a sample of isolated galaxies, suggesting elevated levels of star formation. We suspect that a large fraction of 
nearby early-type spirals are going through an interaction or are in the process of swallowing a smaller galaxy. 
Tentative support for this idea comes from recent HI observations. In a study of 7 Sa-Sab galaxies with high levels of \ha emission, 
\citet{HY2003} find unambiguous signs of recent interactions in 6 of these galaxies. Interaction signatures include 
HI tidal tails (NGC 972, NGC 3885, NGC 7213) 
and neutral hydrogen bridges connecting neighboring galaxies (NGC 3471, NGC 5915, NGC 7582). 

Some of the dispersion seen in the \ha equivalent width vs (B-V) plot of Figure 8 may also be attributed
to interactions. \citet{Kennicutt1987}
find a similar dispersion in their sample of interacting galaxies that can be explained by superimposing a 
burst of star formation on an 
evolved stellar population. The strength of the burst and the location of a galaxy on the plot, of course, 
depends on a number of parameters
including the extinction, the gas content, and the orbital parameters of the interaction. Nevertheless, 
the dispersion of colors and 
\ha equivalent widths seen among early-type spirals is reminiscent of that seen among the
interacting galaxies, providing further support for the idea that 
elevated star formation may only be a temporary phase for these nearby early-type spirals. 

It should be noted, however, that not all interacting early-type spirals exhibit unusually high FIR or \ha emissions. 
For example, HI maps of M81 \citep{Yun1994}, NGC 3368 \citep{Schneider1989}, and 
NGC 4725 \citep{Wevers1984} reveal unambiguous interaction signatures and yet \ha imaging
reveals relatively little evidence for enhanced 
star formation. On the other hand, it is to be expected that the HI signature
of an interaction will last far longer
than the associated episode of massive star formation that the interaction causes.

\subsection{Influence of Bars on Early-type Spirals}
It is now well documented that the presence of a bar in a galaxy can induce star formation 
either at the ends of the bar or in the nuclear and circum-nuclear regions.  The influence of 
bars on global star formation rates has been the subject of many studies over the past few decades 
and has led to the concensus that there are no  
significant differences in {\it global} massive star formation rates between barred and unbarred galaxies
(e.g. \citet{RD1994}; \citet{Tomita1996}). However, there is some evidence that nuclear star formation 
appears to be 
enhanced in early-type barred spirals compared to their 
unbarred counterparts \citep{Devereux1987, Huang1996, HFS1997}. Such an effect is not 
seen in late-type spiral galaxies and may be related to the location of the inner Lindblad resonance (ILR), 
which is dependent 
on the central mass distribution. The ILR in an early-type spiral is expected to be located inside the bar and 
closer to the 
nucleus than in a late-type spiral \citep{EE1985} and thus can transfer gas more efficiently to 
the central 
region of the galaxy, triggering star formation. It should be noted, however, that not all barred early-type 
spirals have 
enhanced nuclear star formation; thus the mere presence of a bar is not sufficient for nuclear 
star formation to occur.

Our sample of 51 Sa-Sab galaxies provides an ideal opportunity to study the effects of bars on these bulge-dominated
galaxies. Our sample includes 29 barred galaxies (including the intermediate types classified as 'X' in the Nearby 
Galaxies Catalog \citep{Tully1988} or 'SAB' in the RC2 catalog \citep{devauc1959}), 17 unbarred galaxies and 
5 galaxies with 
no classification (Table 4). Figure 11 shows the histogram of the ratio of \ha emission from the central 1 kpc 
(diameter) 
region to the total \ha emission for barred and unbarred early-type spirals. We do not find a statistically 
significant difference between barred and unbarred galaxies
in the fraction of the global \ha emission that is radiated by the nucleus. Similarly, figure 12 plots 
the total \ha luminosity against
the ratio of nuclear to total \ha luminosity, and again we find no effect due to the presence of a bar. 
However, the histogram of 
nuclear (1-kpc) \ha luminosity (Figure 13) appears to be slightly enhanced in barred early-type spirals compared to 
unbarred galaxies, consistent
with earlier studies \citep{Devereux1987, Huang1996, HFS1997}.  

Thus, we find some evidence that bars in early-type spirals may enhance the nuclear \ha luminosity, but 
the picture is far from clear. 
We have examples
of barred galaxies with nuclear starbursts (e.g. NGC 3504, NGC 1022), barred galaxies with virtually no star formation 
in the nuclear region (e.g. NGC 1398, NGC 1350), unbarred galaxies with nuclear starbursts (e.g. NGC 3885) 
and an unbarred galaxy, NGC 7213, with a circumnuclear star forming ring.

\section{Conclusions}

We have presented results from an \ha imaging survey of 51 nearby Sa-Sab galaxies. Our images indicate 
that, contrary to popular perception, early-type spirals are, in fact, a diverse group of galaxies 
that span a wide range of massive star formation rates and star formation histories. Some of the 
diversity is attributed to ongoing star formation induced by interactions or mergers. Our 
\ha images also show that giant \hii regions (L(\ha) $\ge$ 10$^{39}~ \lum$)  
do exist in the disks of some early-type spirals thereby dispensing with the need for a special,
different,
or biased initial mass function in early type spirals. Additionally, we have established an essentially one to
one correspondence between the spectroscopically classified LINERS and the existence of an extended and
diffuse component of \ha emission under the bulges of the LINER host galaxies. We also find 
a systematic difference between the morphologies of galaxies in the Revised Shapley 
Ames Catalog and the Second Reference Catalog that may be responsible for 
the contrasting results in the published literature concerning massive star formation rates
among the Hubble sequence of spiral galaxies. 

\acknowledgments
The authors would like to thank Rene' Walterbos for providing the \ha filter set for the APO observations and the staff at 
APO and KPNO for their expert assistance at the telescopes. We would also like to thank the referee for useful comments 
that improved the presentation of the paper.

\clearpage

\begin{figure}
\plotone{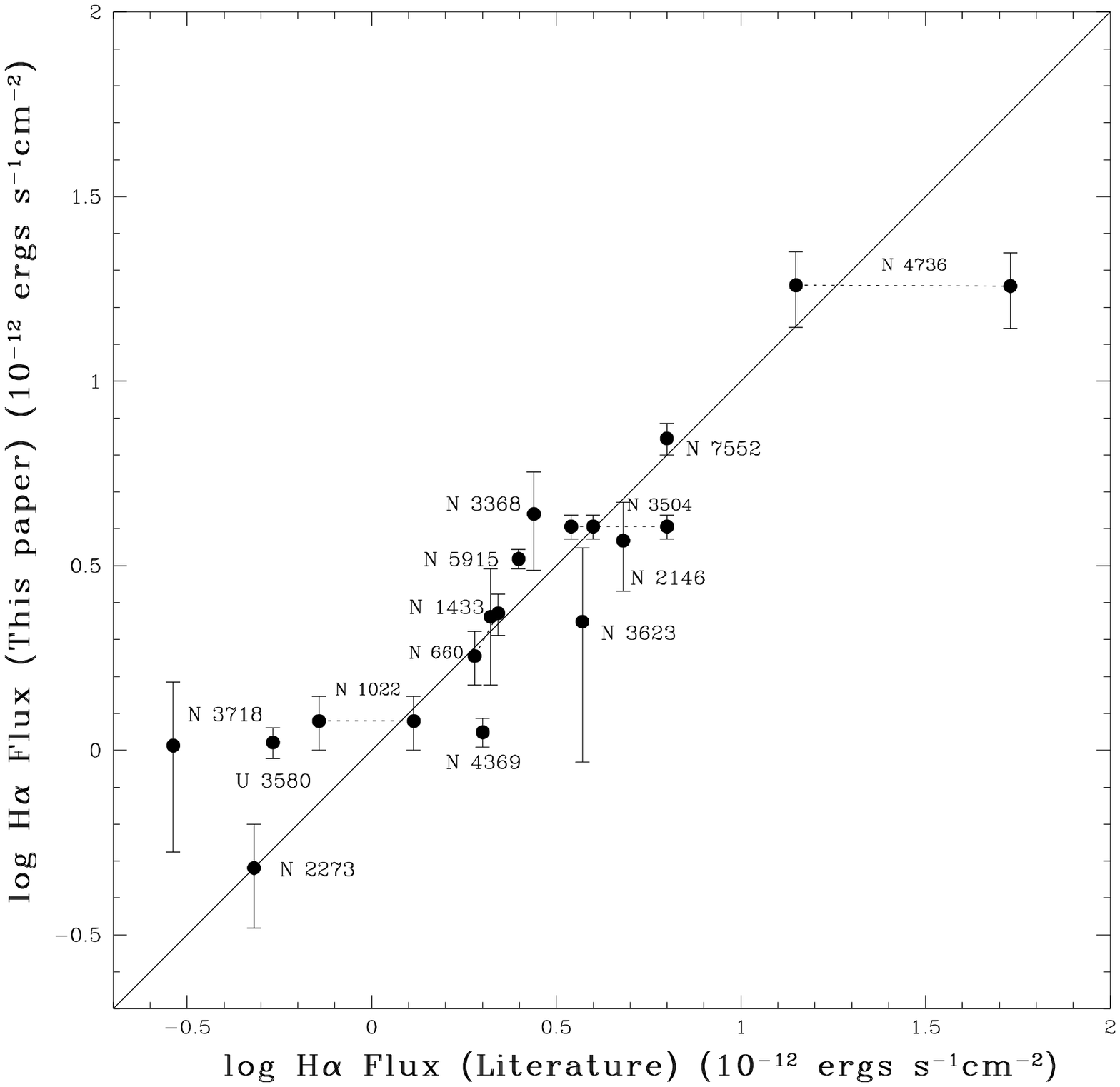}
\caption{Comparison of \ha measurements taken from the literature with those 
measured from our \ha images using the same aperture size (when avialable). NGC 660, 
NGC 1022 and NGC 4736 have two comparison measurements, whereas, NGC 3504 has three published 
values.\label{fig1}}
\end{figure}

\clearpage
\begin{figure}
\caption{a-d: Red continuum and continuum-subtracted \ha images of 12 Category 1 
early-type spiral galaxies. North is at the top and east is at the left in each image.
The white bar in the bottom left-hand corner of each image represents 1 kpc in length.\label{fig2a}}
\end{figure}

\clearpage
\begin{figure}
\caption{a-b: Red continuum and continuum-subtracted \ha images of 6 Category 2 
early-type spiral galaxies. North is at the top and east is at the left in each image.
The white bar in the bottom left-hand corner of each image represents 1 kpc in length.\label{fig3a}}
\end{figure}

\clearpage
\begin{figure}
\plotone{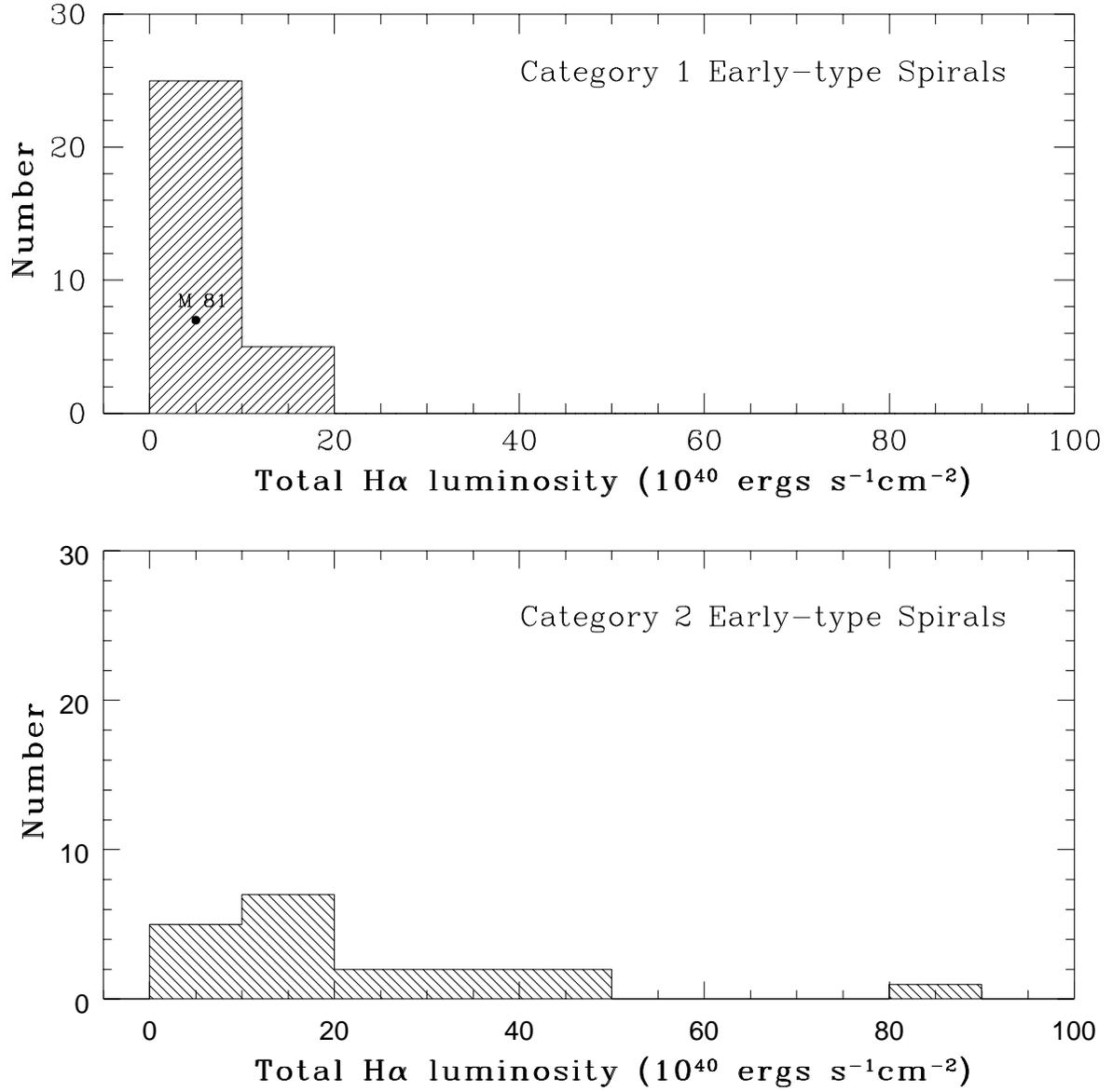}
\caption{Histograms illustrating the distribution of total \ha luminosity of 
Category 1 and Category 2 early-type spirals.\label{fig4}}
\end{figure}

\clearpage
\begin{figure}
\plotone{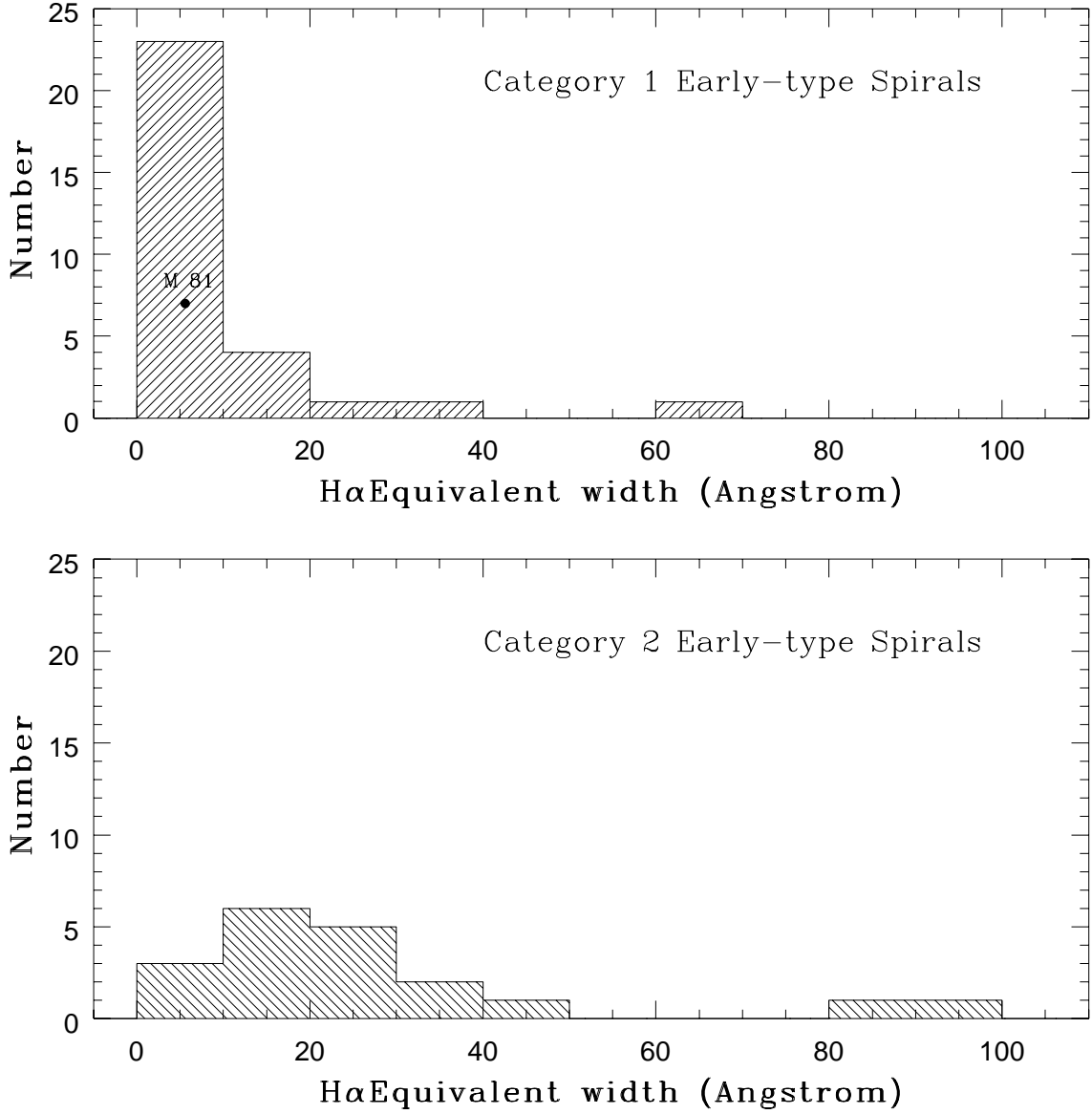}
\caption{Histograms illustrating the distribution of \ha equivalent widths of Category 1 
and Category 2 early-type spirals.\label{fig5}}
\end{figure}

\clearpage
\begin{figure}
\plotone{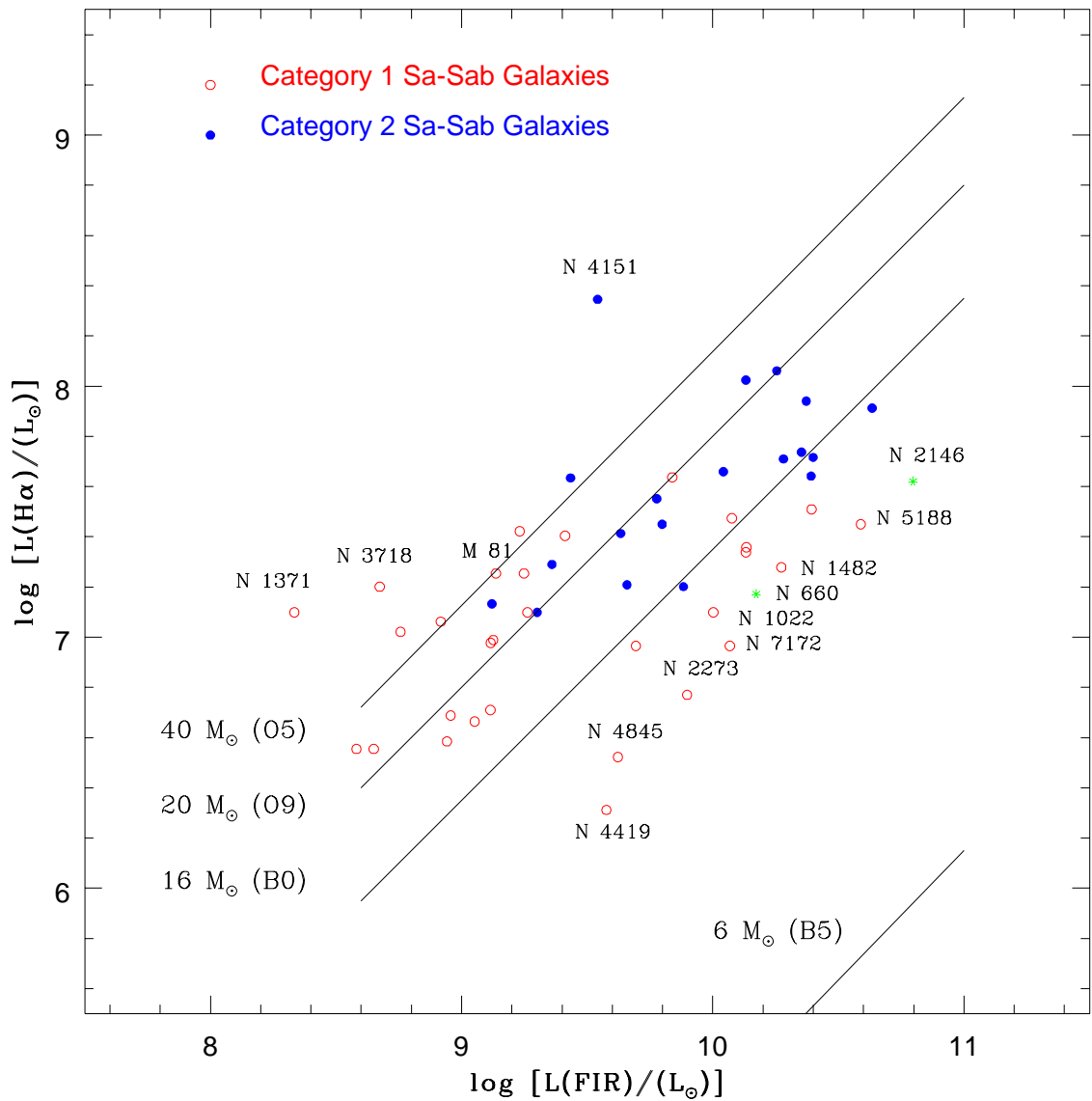}
\caption{Correlation between the \ha luminosity and the FIR (40-120$\mu$m) luminosity
for 51 early-type spirals. The lines represent the $L(FIR/L(H\alpha)$ ratios expected 
for an \hii region powered by stars of masses 40, 20, 16, and 6 M$_{\odot}$, respectively. The graph 
shows that O and B stars are capable of powering both the FIR and \ha luminosities 
in these galaxies. Open circles represent Category 1 galaxies, and solid dots denote Category 2 galaxies. 
A star symbol is used to identify the two unclassified galaxies, NGC 660 and NGC 2146.\label{fig6}}
\end{figure}

\clearpage
\begin{figure}
\plotone{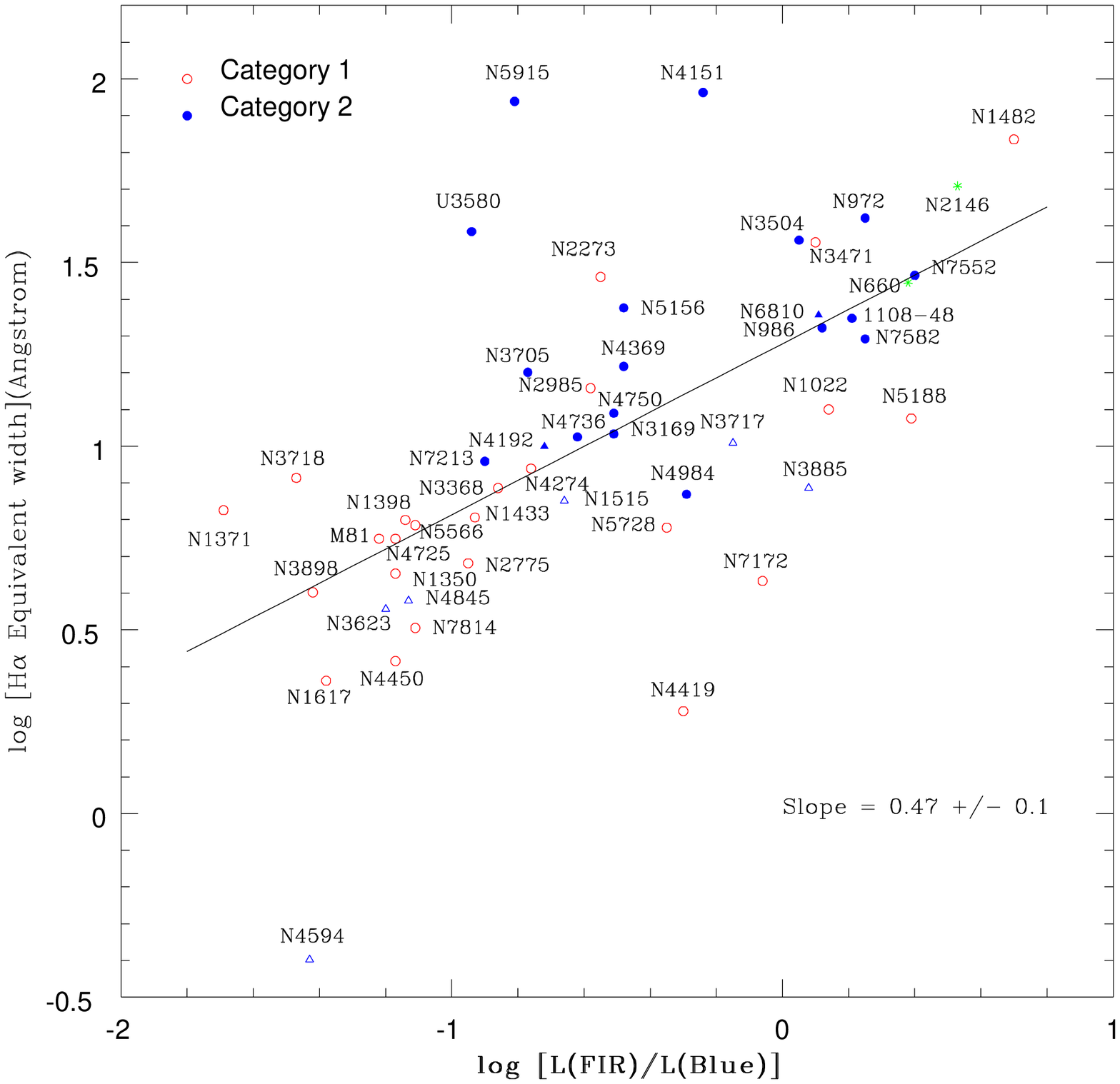}
\caption{Correlation between the \ha equivalent widths and the L(FIR)/L(Blue) luminosity 
ratio for 51 early-type spirals.  Open circles represent Category 1 galaxies and solid dots denote 
Category 2 galaxies. We have also identified highly inclined ($i > 75^\circ$) galaxies in each category as
triangles and we do not find any strong dependence on inclination.  A star symbol is used to 
identify the two unclassified galaxies, NGC 660 and NGC 2146 (NGC 660 also has $i > 75^\circ$). The 
least-squares fit for galaxies is shown as a solid line and all of the galaxies have been identified.\label{fig7}}
\end{figure}

\clearpage
\begin{figure}
\plotone{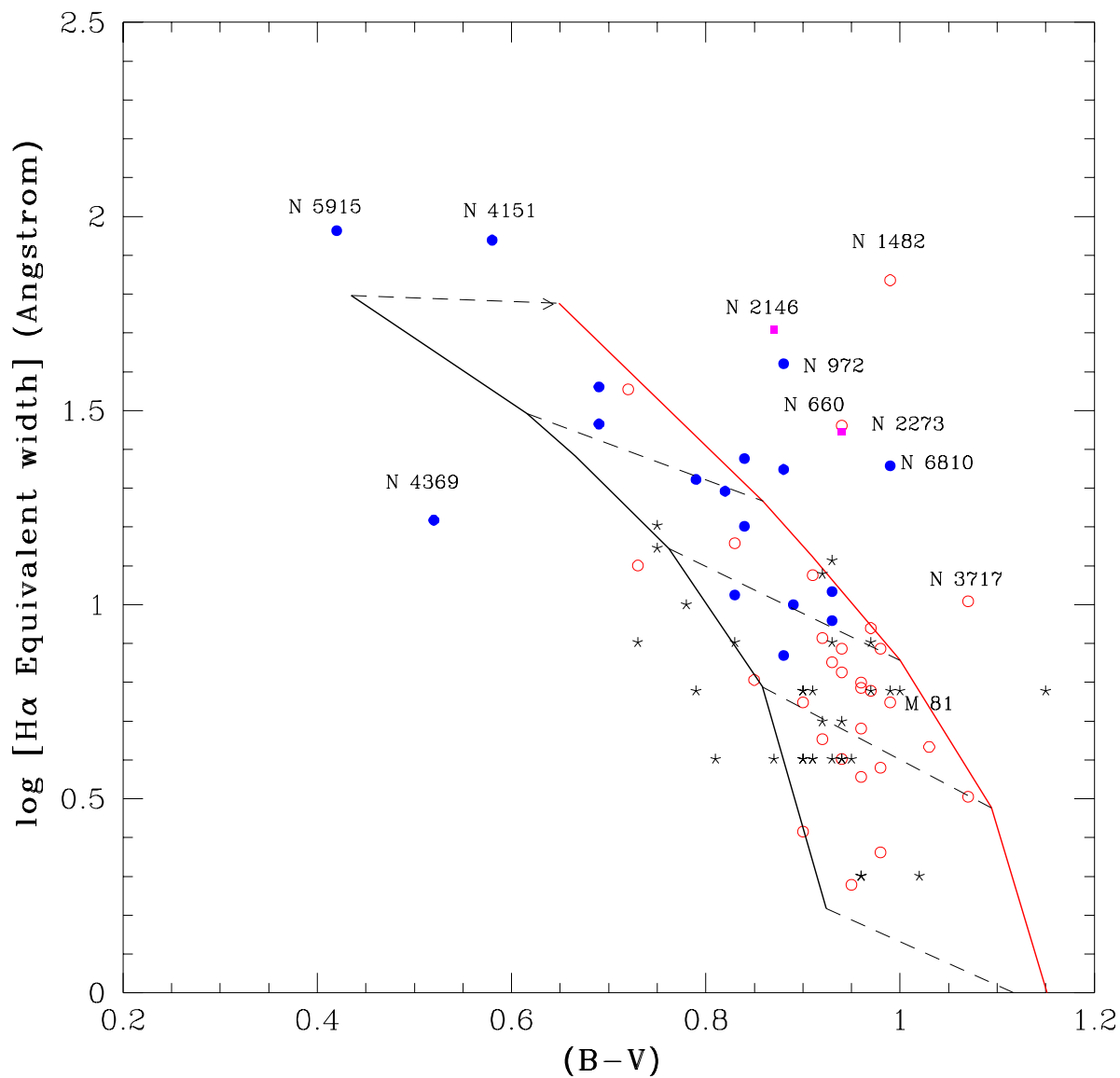}
\caption{Correlation between the (B-V)$_{e}$ colors and \ha equivalent 
widths for 49 early-type spirals. The symbols are the same as Figure 6. The red and 
black lines represent the loci of (B-V)$_{e}$ colors and \ha equivalent widths for 15 Gyr 
old model galaxies that have evolved at different rates, with and without reddening
respectively. The dashed lines indicate the direction of the reddening vectors for 
a constant star formation rate and exponentially decreasing star formation rates 
with with exponential timescales of 10, 5, 3.5 and 2.5 Gyrs.  Kennicutt \& Kent (1983) galaxies
are plotted for comparison as five-pointed stars.\label{fig8}}
\end{figure}

\clearpage
\begin{figure}
\plotone{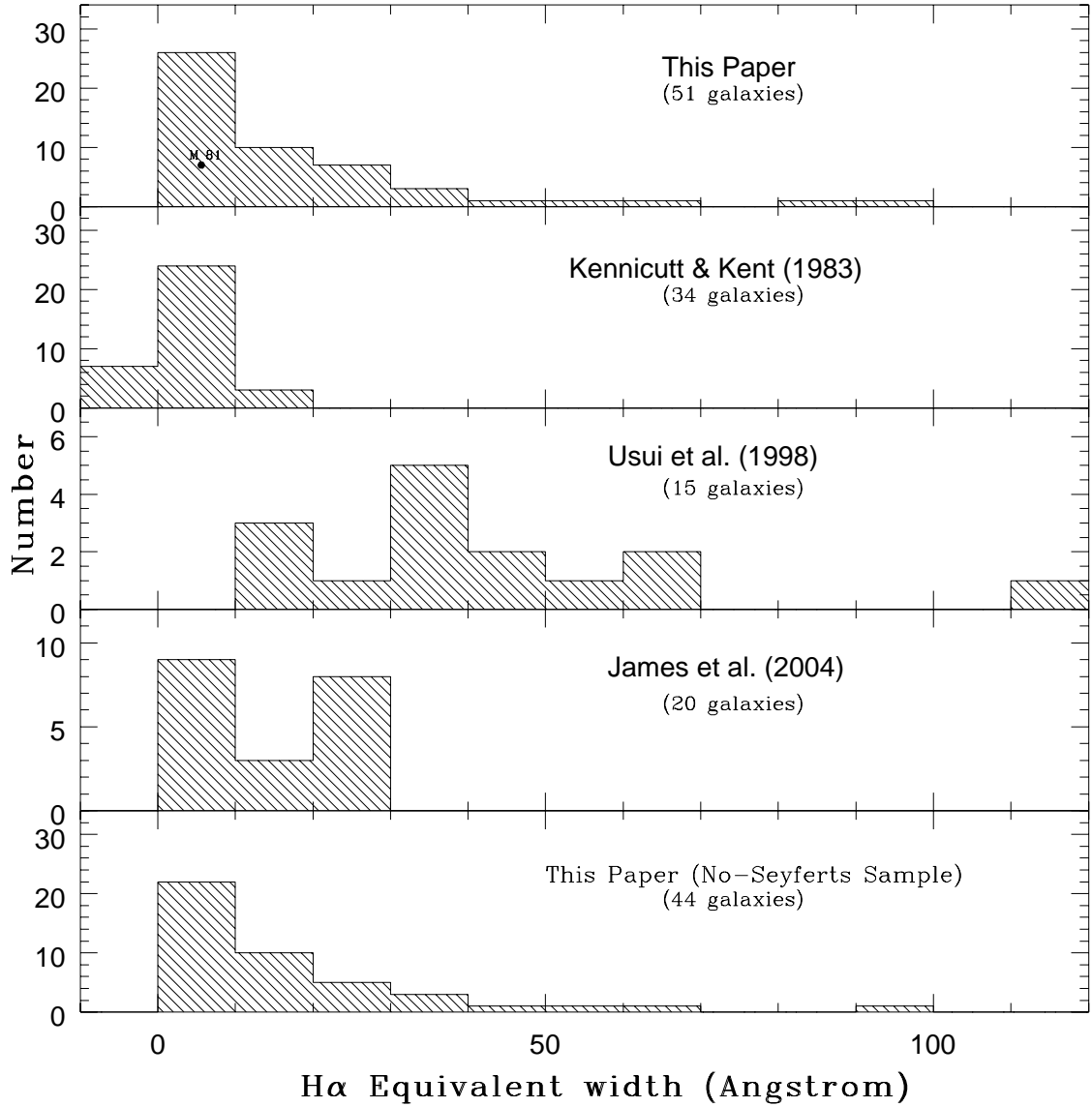}
\caption{Comparison of \ha equivalent widths of all 51 early-type spiral from 
our survey with the \ha equivalent widths of Kennicutt \& Kent (1983), Usui \etal
(1998) and James \etal (2004). Bottom panel is our data after excluding Seyferts.\label{fig9}}
\end{figure}

\clearpage
\begin{figure}
\plotone{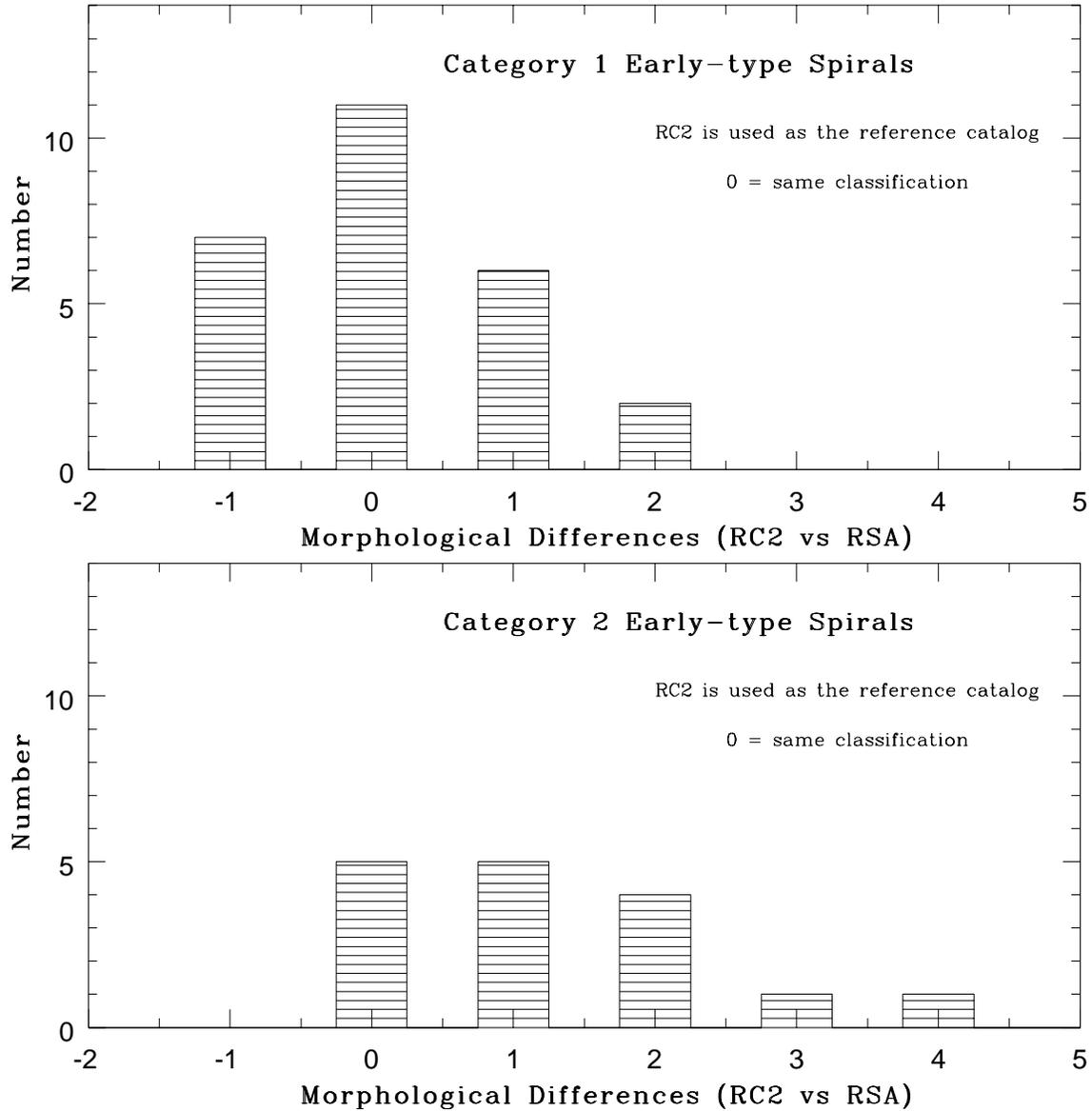}
\caption{Comparison of morphologies from RSA and RC2 catalogs. The RC2 catalog is used as
the reference catalog for Sa-Sab classifications. A ``0''  means that classifications match between 
RC2 and RSA. Differences in classifications for Category 1 galaxies are random. However, 
Category 2 galaxies are preferentially classified as later-types in the RSA catalog. See text for detail.\label{fig10}}
\end{figure}

\clearpage
\begin{figure}
\plotone{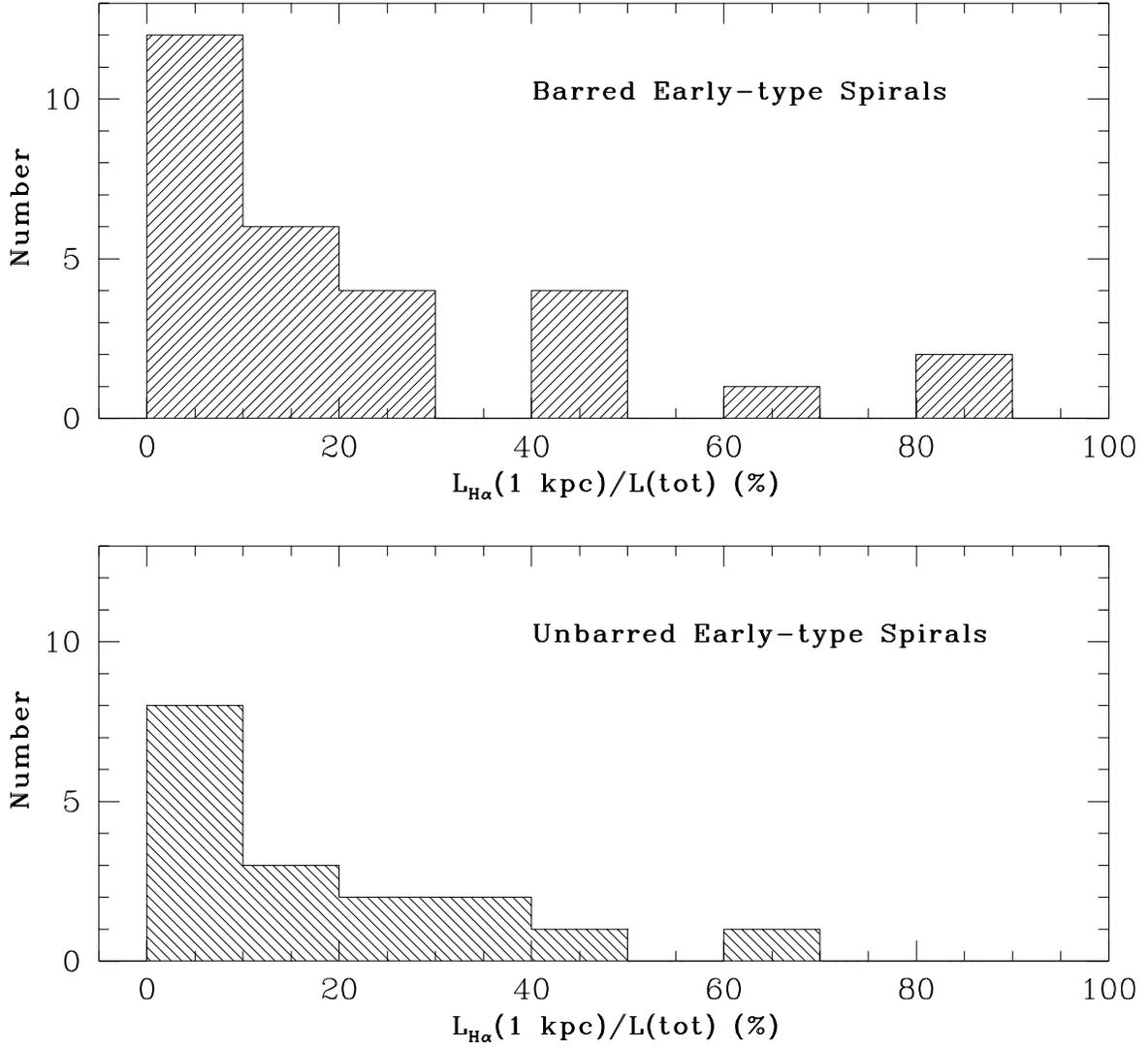}
\caption{Histograms illustrating the ratio of \ha emission from central 1 kpc to total \ha luminosity for barred 
and unbarred  galaxies.\label{fig11}}
\end{figure}

\clearpage
\begin{figure}
\plotone{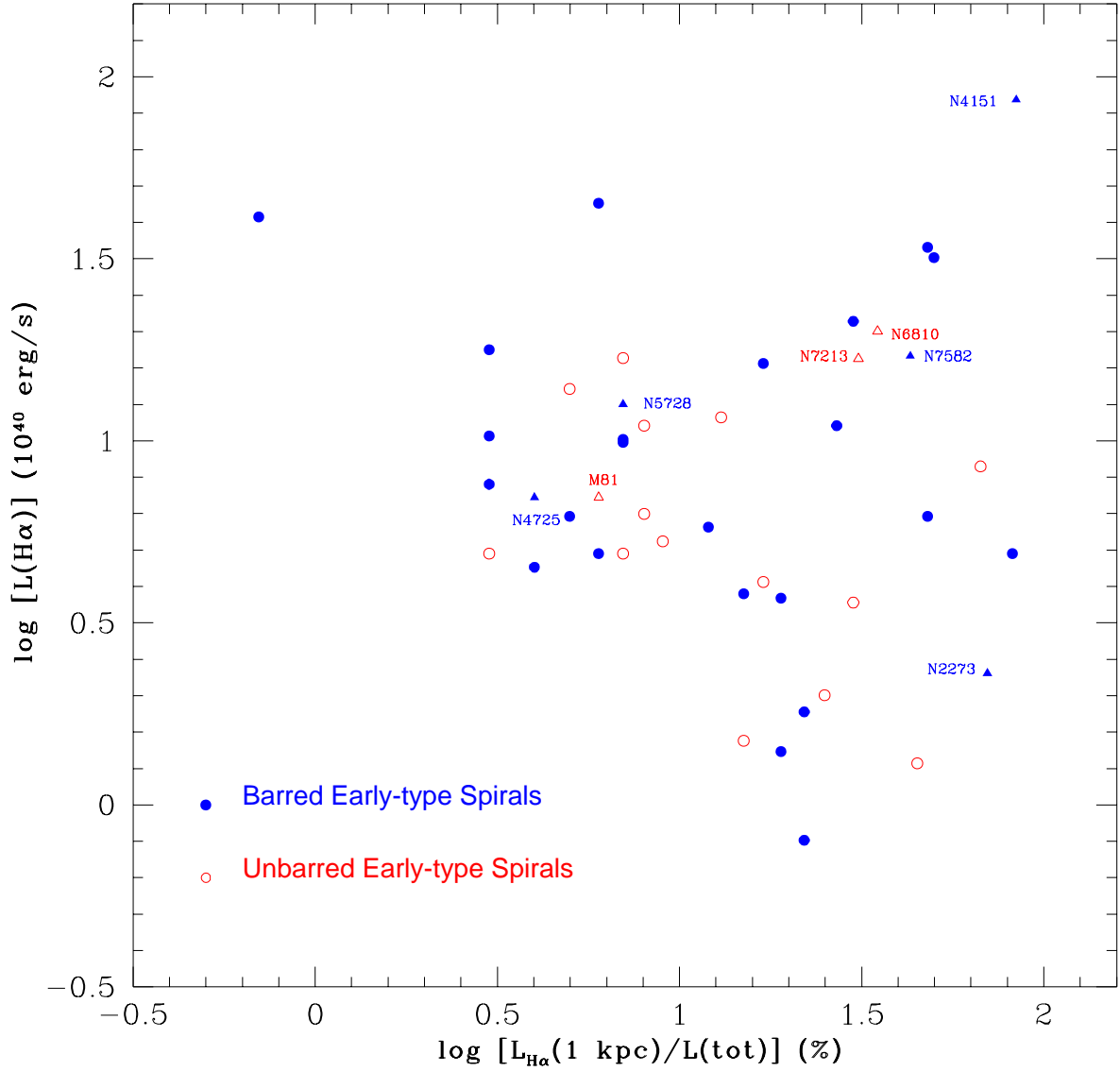}
\caption{Correlation between the ratio of \ha emission from central 1 kpc to total \ha luminosity vs total 
\ha luminosity for barred and unbarred early-type spirals.\label{fig12}}
\end{figure}

\clearpage
\begin{figure}
\plotone{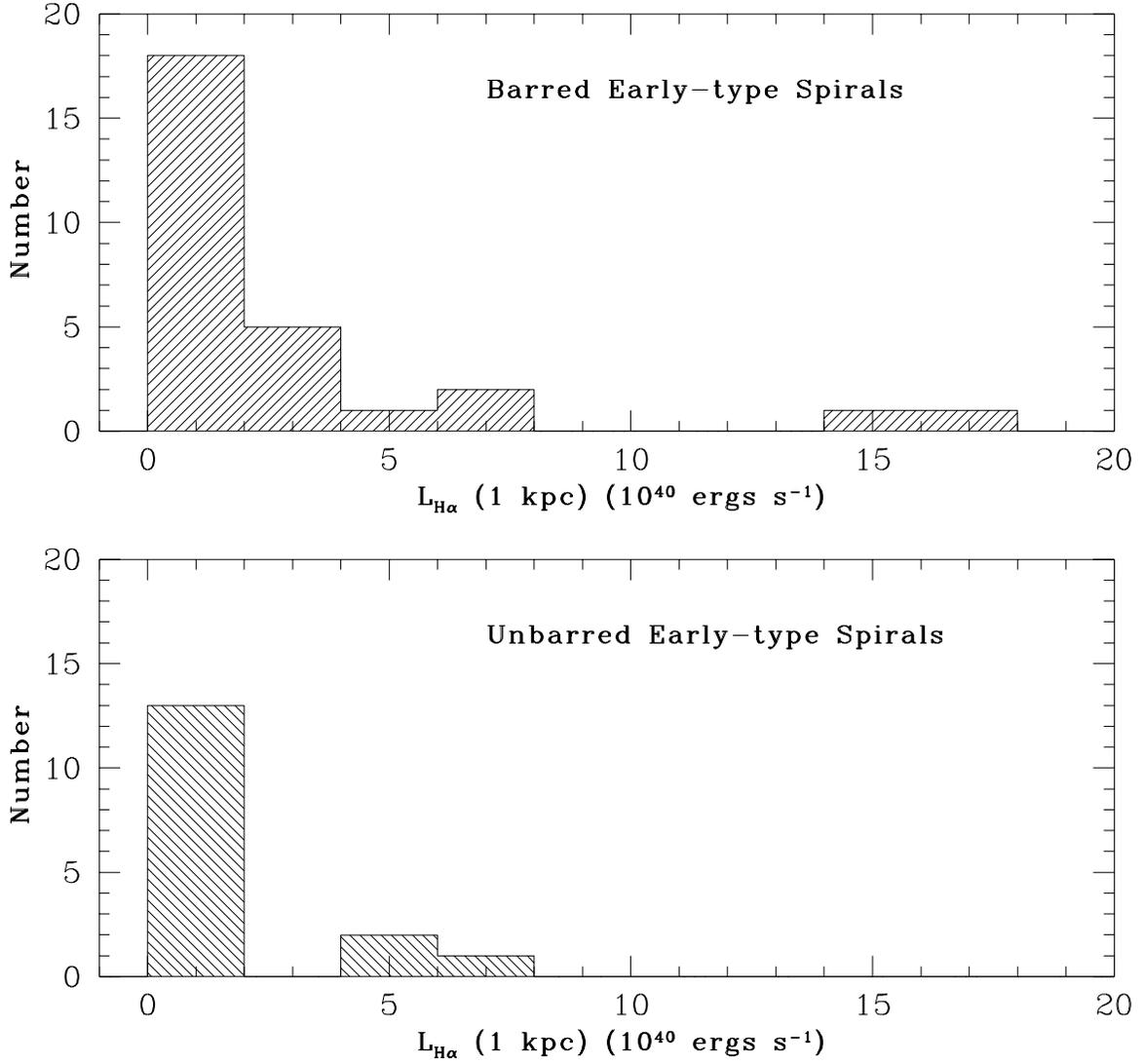}
\caption{Histograms illustrating the \ha luminosity from the central 1 kpc region for barred and unbarred galaxies.\label{fig13}}
\end{figure}






\clearpage

\begin{deluxetable}{lcccccc}
\tabletypesize{\scriptsize}
\tablenum{1}
\tablewidth{0pt}
\tablecaption{Galaxy Parameters\tablenotemark{a}}
\tablehead{
\colhead{Galaxy} & \colhead{m(B)} & \colhead{size} & 
\colhead{$i$} & \colhead{V$_h$} & \colhead{Distance}\\
& \colhead{(mag)} & \colhead{(arc min)} & (deg) & 
\colhead{($kms^{-1}$)} & \colhead{(Mpc)}
}

\startdata
NGC 7814  &11.26   & 5.9 & 68$^\circ$ & 1047 &  15.1  \\
NGC 660  &11.37    & 7.2 & 77$^\circ$ & 856 &  11.8  \\
NGC 972  &11.75    & 3.9 & 65$^\circ$ & 1539 & 21.4 \\
NGC 986  &11.66    & 3.3 & 42$^\circ$ & 1983 & 23.2 \\
NGC 1022 &12.13    & 2.5 & 28$^\circ$ & 1503 & 18.5 \\
NGC 1350 &11.16    & 5.0 & 62$^\circ$ & 1786 & 16.9 \\
NGC 1371 &11.43    & 6.8 & 53$^\circ$ & 1472 & 17.1 \\
NGC 1398 &10.47    & 7.6 & 50$^\circ$ & 1401 & 16.1 \\
NGC 1433 &10.64    & 5.9 & 27$^\circ$ & 1071 & 11.6 \\
NGC 1482 &13.50    & 2.1 & 58$^\circ$ & 1655 &  19.6\\
NGC 1515 &11.17    & 5.7 & 89$^\circ$ & 1169 & 13.4 \\
NGC 1617 &10.92    & 4.0 & 65$^\circ$ & 1040 & 13.4 \\
NGC 2146 &11.00    & 5.3 & 36$^\circ$ & 918  &  17.2\\
NGC 2273 &11.63    & 3.4 & 50$^\circ$ & 1844 &  28.4\\
UGC 3580  &11.90    & 4.0 & 66$^\circ$ & 1204 & 20.6 \\
NGC 2775  &11.09    & 4.6 & 39$^\circ$ & 1135 & 17.0 \\
NGC 2985 &11.18    & 4.0 & 42$^\circ$ & 1277 & 22.4 \\
NGC 3031 &7.59 	   & 22.1 & 60$^\circ$ & -43  & 3.6\tablenotemark{b} \\
NGC 3169 &11.24    & 5.0 & 59$^\circ$ & 1229 &  19.7\\
NGC 3368 &10.05    & 6.7 & 50$^\circ$ & 899 & 8.1 \\
NGC 3471 &12.98    & 1.9 & 64$^\circ$ & 2076 &  33.0\\
NGC 3504 &11.79    & 2.6 & 35$^\circ$ & 1529 & 26.5 \\
1108-48  &13.64    & 2.4 & 53$^\circ$ & 2717 &  35.2\\
NGC 3623 &10.17    & 8.4 & 81$^\circ$ & 806 & 7.3 \\
NGC 3705 &11.31    & 4.6 & 67$^\circ$ & 1017 & 17.0 \\
NGC 3717 &11.87    & 6.1 & 90$^\circ$ & 1731 &  24.6\\
NGC 3718 &11.26    & 9.6 & 66$^\circ$ & 987 & 17.0 \\
NGC 3885 &12.56    & 2.9 & 77$^\circ$ & 1948 &  27.8\\
NGC 3898 &11.71    & 4.7 & 46$^\circ$ & 1172 &  21.9 \\
NGC 4151 &11.13    & 6.3 & 33$^\circ$ & 989 & 20.3 \\
NGC 4192 &10.72    & 8.7 & 83$^\circ$ & -142  &  16.8 \\
NGC 4274 &11.12    & 6.5 & 72$^\circ$ & 922 & 9.7 \\
NGC 4369 &11.80    & 2.4 & 0$^\circ$ & 1052  &  21.6 \\
NGC 4419 &11.93    & 3.0 & 75$^\circ$ & -273  &  16.8 \\
NGC 4450 &10.90    & 5.0 & 50$^\circ$ & 1958  &  16.8 \\
NGC 4594 &9.16    & 8.4 & 79$^\circ$ & 1127&  20.0\\
NGC 4725 &9.91    & 10.5 & 43$^\circ$ & 1207 &  12.4\\
NGC 4750 &11.81    & 2.3& 19$^\circ$ & 1518 &  26.1\\
NGC 4736  &8.85    & 12.2 & 33$^\circ$ & 307 &  4.3\\
NGC 4845 &12.07   & 4.8 & 81$^\circ$ & 1228 &  15.6\\
NGC 4984 &11.68   & 2.3 & 41$^\circ$ & 1259 &  21.3\\
NGC 5156 &11.92    & 2.4 & 24$^\circ$ & 2983 &  39.5\\
NGC 5188 &12.58    & 3.8 & 74$^\circ$ & 2366 &  32.9\\
NGC 5566 &11.29    & 5.6 & 71$^\circ$ & 1518 &  26.4\\
NGC 5728 &11.75    & 2.3 & 65$^\circ$ & 2970 &  42.2\\
NGC 5915 &11.88    & 1.4 & 42$^\circ$ & 2272 &  33.7\\
NGC 6810 &11.40    & 3.2 & 82$^\circ$ & 1995 &  25.3 \\
NGC 7172 &12.55    & 2.1 & 64$^\circ$ & 2651 &  33.9 \\
NGC 7213 &11.35    & 2.1 &  ...   & 1778 & 22.0 \\
NGC 7552 &11.31    & 3.5 & 31$^\circ$ & 1609 & 19.5 \\
NGC 7582 &11.06    & 4.5 & 65$^\circ$ & 1459 & 17.6 \\

\tablenotetext{a}{From \citet{Tully1988}}
\tablenotetext{b}{Distance for NGC 3031 adopted from \citet{Freedman2001}}

\enddata
\end{deluxetable}

\begin{deluxetable}{lccccc}
\tabletypesize{\scriptsize}
\tablenum{2}
\tablewidth{0pt}
\tablecaption{Details of Observations}
\tablehead{
\colhead{Galaxy}	& \colhead{Epoch}	& \colhead{Telescope} &
\colhead{H$\alpha$ filter} & \colhead{Exp. time} & \colhead{Standard\tablenotemark{a}} 
}

\startdata
NGC 7814  &  1999 Jan 13 & KPNO 0.9m & 6586/72 & 1200s &  BD$+28^\circ$4211 \\
UGC 3580  & 1999 Jan 12    & KPNO 0.9m &  6586/72 & 1200s &  BD$+28^\circ$4211 \\
NGC 2775  & 1999 Jan 13/Apr 10    &  KPNO 0.9m & 6586/72 & 900s & BD$+28^\circ$4211 \\
NGC 2985 & 1998 Mar 25    & APO 3.5m & 6610/70 & 390s & PG0934+554  \\
NGC 3368 &  1999 Jan 12   &  KPNO 0.9m & 6586/72 & 1200s &  BD$+28^\circ$4211\\
NGC 3504 & 1999 Feb 13    & APO 3.5m & 6610/70 & 300s & Feige 34  \\
NGC 3623 &  1999 Jan 12   &  KPNO 0.9m & 6586/72  & 1200s & BD$+28^\circ$4211  \\
NGC 3718 & 1999 Apr 10    & KPNO 0.9m & 6586/72  & 1200s &  Feige 34 \\
NGC 3898 &  1999 Apr 12   & KPNO 0.9m &  6586/72 & 1200s &  Feige 34 \\
NGC 4151 &  1999 Apr 13   & KPNO 0.9m & 6586/72  & 1200s & Feige 34 \\
NGC 4274 &   1999 Apr 9  & KPNO 0.9m  & 6586/72  & 1200s &   Feige 34 \\
NGC 4369 & 1999 Feb 14    & APO 3.5m & 6610/70 & 300s & HZ 44 \\
NGC 4594 &  1999 Apr 13     &  KPNO 0.9m & 6586/72 & 1200s &  Feige 34 \\
NGC 4725 &  1999 Apr 10  & KPNO 0.9m  & 6586/72  & 1200s & Feige 34 \\
NGC 4750 & 1999 Feb 13    &  APO 3.5m &  6610/70 & 300s &  Feige 34 \\
NGC 4736 &   1999 Apr 12  &  KPNO 0.9m & 6586/72  & 1200s & Feige 34 \\
NGC 4845 &    1999 Apr 13 &  KPNO 0.9m & 6586/72  & 1200s & Feige 34 \\
NGC 5566 &    1999 Apr 12 & KPNO 0.9m & 6586/72 & 1200s &   Feige 34 \\

\tablenotetext{a}{\citet{Massey1988}}

\enddata
\end{deluxetable}

\begin{deluxetable}{lccccccccc}
\rotate
\tabletypesize{\scriptsize}
\tablecolumns{9}
\tablenum{3}
\tablewidth{0pt}
\tablecaption{Summary of Results}
\tablehead{
\colhead{Galaxy}  &  \colhead{F$_{(H\alpha + N[II])}$\tablenotemark{a}} & \colhead{Aperture}      
& \colhead{L$ _{(H\alpha + N[II])}$}   & \colhead{EW$_{(H\alpha + N[II])}$}    &
\colhead{$L_{FIR(40-120 \mu m)}$} & \colhead{L$_{H\alpha}$(1 kpc)} & \colhead{L$_{H\alpha}$(1 kpc)/L$_{H\alpha}$} & \colhead{Reference\tablenotemark{b}} \\
& \colhead{$(10^{-12}\flux)$} &  \colhead{(arcsec)} & \colhead{$(10^{40}\lum)$}& \colhead{(\AA)} & \colhead{$(10^{10}L_{\odot})$} & \colhead{$(10^{40}\lum)$} & ($\%$) &
}
\startdata

\cutinhead{Category 1 Early-type Spirals}
NGC 2985 & 2.8$\pm$0.5 & 116& 16.9 & 14.4 & 0.69 & 1.1 & 7 & 1 \\
NGC 5728 & 0.6$\pm$0.2 & 75.3 & 12.6 & 6.0 & 2.48  & 0.9 & 7 & 2  \\
NGC 3717 & 1.6$\pm$0.2 & 180.6 & 11.6 & 10.2 & 1.19 & 1.5 & 13 & 2  \\
NGC 5188 & 0.9$\pm$0.1 & 77.4 & 11.0 & 11.9 & 3.88 & 3.0 & 27 & 2 \\
NGC 1398 & 3.3$\pm$1.2 & 215.0 & 10.3 & 6.3 & 0.17 & 0.3 & 3 & 2 \\
NGC 5566 & 1.2$\pm$0.2 & 170 & 9.9 & 6.1 & 0.26 & 0.7 & 7 & 1 \\
NGC 3471 & 0.7$\pm$0.1  & 20.4  & 8.9   & 35.9 & 1.36 & 5.1 & 57 & 2 \\
NGC 3885 & 0.9$\pm$0.1 & 51.6 & 8.5 & 7.7 & 1.36  & 5.7 & 67 & 2 \\
NGC 1482 & 1.6$\pm$0.2 &  48.9  & 7.4 & 68.5 & 1.87  & 3.7 & 50 & 2 \\
NGC 4725 & 3.8$\pm$1.4 & 364 & 7.0  & 5.6 & 0.18 & 0.3 & 4 & 1 \\
NGC 3031 & 45.3$\pm$13.5& ... & 7.0 & 5.6 & 0.14 & 0.4\tablenotemark{b} & 6 & 3 \\
NGC 3718 & 1.8$\pm$0.5 & 374.0 & 6.2 & 8.2 & 0.05 & 0.3 & 5 & 1 \\
NGC 1022 & 1.2$\pm$0.2 & 77.4 & 4.9 & 12.6 & 1.00  & 4.0 & 82 & 2  \\
NGC 1371 & 1.4$\pm$0.5 & 193.5 & 4.9 & 6.7 & 0.02 & 0.3 & 6 & 2 \\
NGC 2775 & 1.4$\pm$0.8 & 170 & 4.9 & 4.8 &  0.18 & 0.1 & 3 & 1 \\
NGC 1350 & 1.3$\pm$0.7 & 176.3 & 4.5 & 4.5 & 0.08 & 0.2 & 4 & 2 \\
NGC 3898 & 0.7$\pm$0.4  & 204  &  4.1  & 4.0 & 0.06 & 0.7 & 17 & 1 \\
NGC 3368 & 4.9$\pm$1.3 & 251 &  3.8& 7.7 & 0.13  & 0.6 & 15 & 1 \\
NGC 1433 & 2.3$\pm$0.8 & 197.8 & 3.7 & 6.4 & 0.13 & 0.7 & 19 & 2 \\
NGC 4594 & 0.8$\pm$1.0 & 128 $\times$ 544 & 3.6 & 0.4 & 0.49 & 1.1 & 30 & 1 \\
NGC 7172 & 0.3$\pm$0.1 & 53.8 & 3.6 & 4.3 & 1.17 & 0.7 & 19 & 2 \\
NGC 2273 & 0.7$\pm$0.2 & 103.7 & 2.3 & 28.9 & 0.79 & 4.5 & 70 & 2 \\
NGC 4450 & 0.6 & 127 &2.0  & 2.6 & 0.13  & 0.5 & 25 & 4 \\
NGC 4274 & 1.7$\pm$0.5 & 177 &1.9  & 8.7 & 0.09 & 0.3 & 18 & 1 \\
NGC 1515 & 0.9$\pm$0.2 & 307 $\times$ 172 & 1.8 & 7.1 & 0.11 & 0.4 & 22 & 2 \\
NGC 7814 & 0.6$\pm$0.2 &  122  & 1.5  & 3.2& 0.09  & 0.2 & 15 &  1 \\
NGC 1617 & 0.6$\pm$0.6 & 111.8 & 1.4 & 2.3 & 0.04  & 0.2 & 14 & 2 \\
NGC 3623 & 2.2$\pm$1.3 & 204 & 1.4  & 3.6 & 0.04& 0.3 & 19 & 1 \\
NGC 4845 & 0.5$\pm$0.2 & 88 & 1.3 & 3.8 & 0.42 & 0.6 & 45 & 1 \\
NGC 4419 & 0.2 &  77  & 0.8 & 1.9 & 0.38  & 0.2 & 22 & 4 \\

\cutinhead{Category 2 Early-type Spirals }
NGC 4151 & 17.5$\pm$0.4 & 245 & 86.5 & 86.9 & 0.35 & 72.5 & 84 & 1 \\
NGC 5915 & 3.3$\pm$0.2 & 43.0 & 44.9 & 91.9 & 1.80 & 2.7 & 6 & 2 \\
NGC 5156 & 2.2$\pm$0.1 & 103.2 & 41.2 & 23.8 & 1.35  & 0.3 & 0.7 & 2 \\
NGC 3504 & 4.0$\pm$0.3 & 90 & 34.0 & 36.4 & 2.36 & 16.2 & 48 & 1 \\
NGC 7552 & 7.0$\pm$0.7 & 116.1 & 31.9 & 29.2 & 4.31 & 16.0 & 50 & 2 \\
NGC 986 & 3.3$\pm$0.4 & 114.0 & 21.3 & 21.0 & 2.25 & 6.3 & 30 & 2 \\
NGC 972 & 3.7$\pm$0.2 &  58.6  & 20.3 & 41.8 & 2.51  & 1.8 & 9 & 2 \\
NGC 6810 & 2.6$\pm$0.2 & 86.0 & 20.0 & 22.8 & 1.91 & 6.9 & 35 & 2 \\
1108-48 & 1.2$\pm$0.1 & 137.6 & 17.8 & 22.3 & 1.10 & 0.6 & 3 & 2 \\
NGC 7582 & 4.6$\pm$0.7 & 139.8 & 17.1 & 19.6 & 2.46  & 7.4 & 43 & 2 \\
NGC 7213 & 2.9$\pm$0.4 &  210.7  & 16.8  & 9.1 & 0.27 & 5.2 & 31 & 2 \\
NGC 3169 & 3.0$\pm$0.5 & 217.2 & 13.9 & 10.8 & 0.60 & 0.7 & 5 & 2 \\
NGC 4750 & 1.4$\pm$0.3 & 72 & 11.0 & 12.3 & 0.63 & 0.8 & 8 & 1 \\
NGC 4192 & 3.0 &  380 $\times$ 380  & 10.1 & 10.0 & 0.43  & 0.7 & 7 & 4 \\
NGC 3705 & 2.2 & 204 & 7.6 & 15.9 & 0.23 & 0.2 & 3 & 4 \\
NGC 4369 & 1.1$\pm$0.1 & 33 & 6.3 & 16.5 & 0.46 & 0.5 & 8 & 1 \\
NGC 4984 & 1.1 & 80  & 6.2 & 7.4 & 0.76 & 2.9 & 48 & 4 \\
UGC 3580 & 1.1$\pm$0.1 & 136 & 5.3 & 38.4 & 0.13 & 0.5 & 9 & 1 \\
NGC 4736 & 22.1$\pm$4.2 & 476 &  4.9 & 10.6 & 0.20 & 0.3 & 7 & 1 \\

\cutinhead{Unclassified}
NGC 2146 & 4.6$\pm$1.0 & 119.0  & 16.3 & 51.1 & 6.28 & 2.8 & 17 & 2 \\
NGC 660\tablenotemark{c} & 3.5$\pm$0.3 & 173 $\times$ 585 & 5.8 & 27.9 & 1.49 & 0.7 & 12 & 1 \\

\enddata
\tablenotetext
{a}{Fluxes have not been corrected for Galactic or internal extinction. \ha+[NII] fluxes 
for galaxies observed by \citet{Koopman1997} have 20-30$\%$ uncertainty.} 
\tablenotetext
{b}{Nuclear H$\alpha$ luminosity for NGC 3031 is measured for central 
800 $\times$ 800 pc$^{2}$ region \citep{Devereux1995}}
\tablenotetext
{c}{NGC 660 was also included in \citet{HD1999}. However, these are new large field of view \ha observations 
that include the full extent of the galaxy.}
\tablerefs{(1) This Paper; (2) \citet{HD1999}; (3) \citet{Devereux1995}; (4) \citet{Koopman1997}}
\end{deluxetable}

\begin{deluxetable}{lccccl}
\rotate
\tabletypesize{\scriptsize}
\tablecolumns{6}
\tablenum{4}
\tablewidth{0pt}
\tablecaption{Summary of Results II}
\tablehead{
\colhead{Galaxy}           &  \colhead{Nuc. morph.\tablenotemark{a}}      &
\colhead{Nuc. Sp. class\tablenotemark{b}}          & \colhead{Ref.}  &
\colhead{Bar\tablenotemark{c}}        &  \colhead{Comments} 
}
\startdata
\cutinhead{Category 1 Early-type Spirals }
NGC 2985 & PS ?  & T1.9 & 2 & N & \\
NGC 5566 & ENER & L2 & 2 & Y & Disturbed \ha morphology. \\
NGC 5728 & PS & S2 & 3 & X &  A tightly wound inner spiral arm and a loose outer spiral
arm in the continuum.\\
NGC 3717 & ? & H & 4 & N &  A prominent dust lane parallel to the major axis of the galaxy\\
NGC 5188 & PS & H & 4 & Y &  \\
NGC 1398 & ENER & N & 5 & Y &  A bar-ring morphology in \ha.\\
NGC 3471 & SB & H  & 6 & ? &  \\
NGC 3885 & SB & H & 7 & N & \\
NGC 3718 & ? & L1.9 & 2 & Y & Disturbed continuum morphology.\\
NGC 1482 & SB & \nodata & \nodata & ? & A prominent dust lane parallel to the major axis of the galaxy. \\
& & & & & Filaments and/or chimneys of ionized gas extending perpendicular to the disk.\\
NGC 4725 & ENER & S2 & 2 & X & \\
NGC 3031 & ENER/PS & S1 & 2 & N & \\
NGC 1022 & SB & H & 8 & Y & A tightly wound inner spiral arm and a loose outer spiral
arm in the continuum.\\
NGC 1371 & ENER & \nodata & \nodata & X & \\
NGC 2775 & ENER & \nodata & \nodata & N & \\
NGC 1350 & ENER & N & 4 & Y &  A bar-ring morphology in \ha.\\
& & & & &  A tightly wound inner spiral arm and a loose outer spiral arm in the continuum.\\
NGC 3898 & ENER & T2 & 2 & N & \\
NGC 3368 & ENER &  L2 & 2 & X & A bar-ring morphology in \ha.\\
NGC 1433 & ENER & N & 4 & Y & A bar-ring morphology in \ha. \\
& & & & & A tightly wound inner spiral arm and a loose outer spiral
arm in the continuum.\\
NGC 4594 & ? & L2 & 2 & N & \\
NGC 7172 & PS & S2 & 8 & ? & A prominent dust lane parallel to the major axis of the galaxy \\
NGC 2273 & PS & S2 & 2 & Y & \\
NGC 4450 & ENER & L1.9 & 2 & N & \\
NGC 4274 & ? & H &  2 & Y & \\
NGC 1515 & ENER & \nodata & \nodata & X & \\
NGC 7814 & ? & \nodata & \nodata & N & \\
NGC 1617 & ENER & \nodata & \nodata & ? & \\
NGC 3623 & ENER & L2 & 2 & X & \\
NGC 4845 & ? & H & 2 & N & \\
NGC 4419 & ? & T2 & 2 & Y & \\
\cutinhead{Category 2 Early-type Spirals }
NGC 4151 & \nodata & S1.5 & 2 & X & \\
NGC 5915 & \nodata & \nodata & \nodata & Y & Asymmetric spiral arms in the continuum. \\
& & & & & \ha morphology does not correspond with the major continuum features \\
NGC 5156 & \nodata & \nodata & \nodata & Y &  \\
NGC 3504 & \nodata & H & 2 & X & \\
NGC 7552 & \nodata & H & 4 & Y & A dwarf galaxy at the end of the northern spiral arm \\
NGC 986 & \nodata & H & 4 & Y &  A tidally disrupted dwarf galaxy at the end of the northern arm?\\
NGC 972 & \nodata & H & 2 & ? & Dusty morphology. A possible bar-like structure crossing the nucleus \\
NGC 6810 & \nodata & S2 & 4 & N & A prominent dust lane parallel to the major axis of the galaxy \\
1108-48 & \nodata & \nodata & \nodata & Y & A faint tidal tail leading to a star forming region 18 kpc north-east
of the nuc. \\
NGC 7582 & \nodata & S2 & 9 & Y & A prominent dust lane parallel to the major axis of the galaxy \\
NGC 7213 & PS/ENER &  S1 & 1 & N & A giant \ha filament is located approximately 17 kpc 
south of the galaxy with no \\
& & & & & counterpart in the continuum image.\\
NGC 3169 & ENER & L2 & 2 & N  & \\
NGC 4750 & \nodata & L1.9 & 2 & N & \\
NGC 4192 & \nodata & T2 & 2 & X & \\
NGC 3705 & \nodata & T2 & 2 & X & \\
NGC 4369 & \nodata & H & 2 & N & \ha emission concentrated only near the nuclear region.\\
NGC 4984 & \nodata & \nodata & \nodata & X & \\
UGC 3580 & \nodata &  \nodata & \nodata & N & \\
NGC 4736 & \nodata & L2 & 2 & N & An anomalous \ha tidal arm.\\
\cutinhead{Unclassified}
NGC 2146 & \nodata & H & 2 & Y & Highly disturbed morphology with a prominent dust lane \\
NGC 660 & \nodata & T2/H & 2 & Y &  Highly disturbed morph.Two prominent dust lanes perpendicular to each other\\
\enddata

\tablenotetext
{a}{\ha nuclear morphology: PS=Unresolved point source, ENER=Extended
Nuclear Emission Line Region, SB=nuclear starburst. Nuclear regions of all category 2 and unclassified galaxies are resolved in \ha, 
but they do not meet the starburst or the ENER criteria as defined in the text. Hence 
their nuclear regions have not been classified morphologically.}

\tablenotetext{b}{Classification of the nuclear spectrum: S=Seyfert nucleus,
H=HII nucleus, L=LINER, T=Transition objects, N=Seyfert-like, i.e. H$\alpha < 
1.2 \times [NII] \lambda 6583$}

\tablenotetext
{c}{Bar classifications are taken from \citet{Tully1988} : Y=Bar is present, N=Bar is absent,
X=intermediate case, ?=No information.}

\tablerefs{
(1)\citet{FH1984}; (2)\citet{HFS1997};
(3)\citet{PCB1983}; (4)\citet{VV1986}; (5)\citet{Balzano1983}; 
(6)\citet{LH1995}; (7)\citet{Ashby1995}; (8)\citet{Sharples1984}; (9)\citet{Unger1987} 
}

\end{deluxetable}

\begin{deluxetable}{lcccc}
\tabletypesize{\scriptsize}
\tablenum{5}
\tablecolumns{4}
\tablewidth{0pt}
\tablecaption{Morphological Classifications}
\tablehead{
\colhead{Galaxy}           &  \colhead{NBG\tablenotemark{a}} & \colhead{RC2\tablenotemark{b}}      
& \colhead{RSA\tablenotemark{c}} 
 }
\startdata
\cutinhead{Category 1 Early-type Spirals }
NGC 2985 & Sab   &  Sab   & Sab \\
NGC 5566 & Sab  &  Sab  &  Sa  \\
NGC 5728 & Sa & Sa & Sb  \\
NGC 3717 & Sab & Sb & Sb  \\
NGC 5188 & Sa & Sb & Sbc  \\
NGC 1398 & Sab & Sab & Sab  \\
NGC 3471 & Sa & Sa & \nodata  \\
NGC 3885 & Sa & S0/a & Sa \\
NGC 3718 & Sa   &   Sa    &  Sa \\
NGC 1482 & Sa & S0/a & \nodata  \\
NGC 4725 & Sab   &  Sab  & Sb \\
NGC 3031 & Sab & Sab & Sb \\
NGC 1022 & Sa & Sa & Sa  \\
NGC 2775 & Sab  & Sab  & Sa \\
NGC 1371 & Sa & Sa & Sa  \\
NGC 1350 & Sab & Sab & Sa \\
NGC 3898 & Sab  &  Sab   &  Sa \\
NGC 3368 & Sab  &  Sab   & Sab \\
NGC 1433 & Sa & Sa & Sb \\
NGC 4594 & Sa  &  Sa  &  Sa/Sb \\
NGC 7172 & Sab & Sab & \nodata \\
NGC 2273 & Sa & S0/a & \nodata  \\
NGC 4450 & Sab & Sab & Sab \\
NGC 4274 & Sab  & Sab  &  Sa \\
NGC 1515 & Sa & Sbc & Sb \\
NGC 7814 & Sab &  Sab  &  Sab \\
NGC 1617 & Sa & Sa & Sa  \\
NGC 3623 & Sa  &  Sa  &  Sa \\
NGC 4845 & Sa  &  Sab  &  Sa  \\
NGC 4419 & Sa & Sa & Sab \\
\cutinhead{Category 2 Early-type Spirals }
NGC 4151 &  Sab   & Sab   &  Sab \\
NGC 5915 & Sab & Sab & Sbc \\
NGC 5156 & Sa & Sab & Sbc  \\
NGC 3504 & Sab   & Sab      & Sb \\
NGC 7552 & Sab & Sab & Sbc\\
NGC 986 & Sab & Sab & Sb \\
NGC 972 & Sab & S0/a & Sb  \\
NGC 6810 & Sa & Sab & Sb\\
1108-48 & Sab & \nodata & \nodata  \\
NGC 7582 & Sab & Sab & Sab \\
NGC 7213 & Sa & Sa & Sa \\
NGC 3169 & Sa & Sa & Sb  \\
NGC 4750 & Sab   &  Sab   & Sb \\
NGC 4192 & Sab & Sab & Sb \\
NGC 3705 & Sab & Sab & Sab \\
NGC 4369 & Sa   &  Sa   & Sc \\
NGC 4984 & Sa & \nodata & Sa \\
UGC 3580 & Sa   & \nodata  & \nodata \\
NGC 4736 & Sab  &  Sab  &  Sab \\
\enddata

\tablenotetext
{a}{{\it Nearby Galaxies Catalog}, \citet{Tully1988}} 
\tablenotetext
{b}{{\it Second Ref. Catalogue of Bright Galaxies}, \citet{devauc1976}}
\tablenotetext
{c}{{\it A Revised Shapley-Ames Catalog of Bright Galaxies}, \citet{ST1981}}

\end{deluxetable}

\end{document}